\documentstyle[preprint,aps]{revtex}
\begin{document}
\draft

\title
{Dirac's Observables for the Higgs Model: II) the non-Abelian SU(2) Case.}

\author{Luca Lusanna}

\address
{Sezione INFN di Firenze\\
L.go E.Fermi 2 (Arcetri)\\
50125 Firenze, Italy\\
E-mail LUSANNA@FI.INFN.IT}

\author{and}

\author{Paolo Valtancoli}

\address
{Dipartimento di Fisica\\
Universita' di Firenze\\
L.go E.Fermi 2 (Arcetri)\\
50125 Firenze, Italy\\
E-mail VALTANCOLI@FI.INFN.IT}

\maketitle

\begin{abstract}

We search a canonical basis of Dirac's observables for the classical
non-Abelian Higgs model with fermions in the case of a trivial SU(2)
principal bundle with a complex doublet of Higgs fields and with the
fermions in a given representation of SU(2). Since each one of the three
Gauss law first class constraints can be solved either in the corresponding
longitudinal electric field or in the corresponding Higgs momentum, we
get a priori eight disjoint phases of solutions of the model. The only two
phases with SU(2) covariance are the SU(2) phase with massless SU(2)
fields and the Higgs phase with massive SU(2) fields. The Dirac
observables and the reduced physical (local) Hamiltonian and (nonlocal)
Lagrangian of the Higgs phase are evaluated: the main result is the 
nonanalyticity in the SU(2) coupling constant, or equivalently in the
sum of the residual Higgs field and of
the mass of the SU(2) fields. Some comments on the function spaces needed for
the gauge fields are made.

\vskip 1truecm
\noindent February 1996
\vskip 1truecm
\noindent This work has been partially supported by the network ``Constrained 
Dynamical Systems" of the E.U. Programme ``Human Capital and Mobility".

\end{abstract}
\vfill\eject

\section
{Introduction}

In a previous paper \cite{lv} (quoted as I), we evaluated the Dirac observables
for the Abelian Higgs model\cite{eb} 
with fermions, in the case of a trivial U(1)
principal bundle over a fixed $x^o$ slice of
Minkowski spacetime, for both its electromagnetic and
Higgs phases. Here we will study the simplest non-Abelian Higgs model
(see for instance Ref.\cite{cl}) with fermions, namely the one associated with
a trivial SU(2) principal bundle over a fixed $x^o$ slice of
Minkowski spacetime with a complex 
doublet of Higgs fields and a set of fermion fields belonging to a
representation $\rho$ of SU(2). However, the same construction applies to every 
trivial principal G-bundle with a compact, semisimple, connected, simply
connected structure Lie group G with suitable Higgs fields. As shown in
Ref.\cite{lus}, in this case one can find the Dirac observables of
classical Yang-Mills theory with fermions, if the Yang-Mills gauge
potentials belong to a suitable weigthed Sobolev space in which the Gribov
ambiguity is absent. See Ref.\cite{lus1} for a review of the general
methodology for finding Dirac's observables of physical gauge systems.

\section
{The Lagrangian formalism}

By using the notations of Ref.\cite{lus},
the SU(2) Higgs model is described by the following Lagrangian density
[$\lambda > 0$, $\phi_o > 0$]

\begin{eqnarray}
{\cal L}(x)&=&-{1\over {4g^2}}F_{a\mu\nu}(x)F_a^{\mu\nu}(x)+[D^{(A)}_{\mu}
\phi (x)]^{\dagger}\, D^{(A)\mu} \phi (x) -V(\phi )+
\nonumber \\
&&+{i\over2}\bar \psi (x)[\gamma^{\mu}D_{\mu}^{(A^{(\rho )})}-
\loarrow{D_{\mu}^{(A^{(\rho )}){\dagger}}}\gamma^{\mu}]\psi (x)
- m\bar \psi (x)\psi (x),
\label{1}
\end{eqnarray}

\noindent where the Higgs field $\phi (x)$ is a doublet of complex scalar
fields [$D^{(A){\dagger}}_{\mu}$ means transpose conjugate, not adjoint]

\begin{eqnarray}
&&\phi (x)=(\phi_i(x))=\left( \begin{array}{ll} \phi_1(x)\\ \phi_2(x) 
\end{array} \right),\quad \quad \phi^{\dagger}(x)=(\, \phi^{*}_1(x)\,\,
\phi^{*}_2(x)\, ) \nonumber \\
&&(D^{(A)}_{\mu}\phi (x))_i=(\partial_{\mu}+{\tilde A}_{\mu}(x))_{ij}\phi_j
(x)\nonumber \\
&&{[D_{\mu}^{(A)}\phi (x)]}_i^{\dagger}=\phi^{\dagger}_j(x)
\loarrow{(\partial_{\mu}-{\tilde A}_{\mu}(x))_{ji}}
\nonumber \\
&&{\tilde A}_{\mu}(x)=A_{a\mu}(x){\tilde T}^a, \quad\quad {\tilde T}^a=-i
{{\tau^a}\over 2},\, a=1,2,3,\nonumber \\
V(\phi )&=&\lambda {[\phi^{\dagger}(x)\phi (x)-\phi^2_o]}^2=\mu^2
\phi^{\dagger}(x)\phi
(x)+\lambda [\phi^{\dagger}(x)\phi (x)]^2+\lambda \phi_o^4=\nonumber \\
&=&-{1\over 2}m^2_H\phi^{\dagger}(x)\phi (x)+\lambda [\phi^{\dagger}(x)
\phi (x)]^2+\lambda \phi_o^2,\nonumber \\
&&{}\nonumber \\
&&\mu^2=-2\lambda \phi_o^2 < 0,
\quad\quad m^2_H=-2\mu^2=4\lambda \phi_o^4,\quad\quad
\phi_o={{m_H}\over {2\sqrt{\lambda}}}=\sqrt{ {{-\mu^2}\over {2\lambda}} }
\label{2}
\end{eqnarray}

\noindent The SU(2) generators are
${\tilde T}^a=-{i\over 2}\tau^a=-{({\tilde T}^a)}^{\dagger}$ [$\vec \tau$
are the Pauli matrices]. We take
$\mu^2 < 0$, so that the potential $V(\phi )$ has a set of absolute
minima for $\phi^{\dagger}\phi =\phi_o^2$, parametrized by three  phases 
[see later on Eq.(\ref{16})], and 
$\phi_o > 0$, an arbitrary real number [$< \phi > = \phi_o\not= 0$
at the quantum level: this is the gauge not-invariant formulation
of the statement of symmetry breaking]. 

The gauge potentials $A_{\mu}(x)=A_{a\mu}(x){\hat T}^a=g{\check A}_{\mu}(x)$
[$g$ is the SU(2) coupling constant] belong to the adjoint representation of 
SU(2), whose Lie algebra su(2) has the structure constants $c_{abc}=\epsilon
_{abc}$; we have $({\hat T}^a)_{bc}=\epsilon_{abc}$ and ${\hat T}^a=-
{({\hat T}^a)}^{\dagger}$. The associated field strengths and covariant
derivatives are

\begin{eqnarray}
&&F_{a\mu\nu}(x)=\partial_{\mu}A_{a\nu}(x)-\partial_{\nu}A_{a\mu}(x)+c_{abc}
A_{b\mu}(x)A_{c\nu}(x), \quad F_{\mu\nu}(x)=F_{a\mu\nu}(x){\hat T}^a\nonumber \\
&&{}\nonumber \\
&&{\hat D}^{(A)}_{\mu ab}=\delta_{ab}\partial_{\mu}+c_{abc}A_{b\mu}(x),
\label{3}
\end{eqnarray}

\noindent while the Bianchi identities are ${\hat D}^{(A)}_{\mu}{*}F^{\mu\nu}
\equiv 0$ (${*}F^{\mu\nu}={1\over 2}\epsilon^{\mu\nu\alpha\beta}F_{\alpha\beta}$
).

The fermion fields $\psi (x)=(\psi_{A\alpha}(x))$  [$\alpha$ are spinor indices]
are Grassmann-valued and belong to a representation $\rho$ of SU(2) with
generators $T^a=-{(T^a)}^{\dagger}$; the associated covariant derivative is

\begin{eqnarray}
&&D_{\mu}^{(A^{(\rho )})}\psi (x)=(\partial_{\mu}+A^{(\rho )}_{\mu}(x))\psi (x)
\nonumber \\
&&\bar \psi (x)\loarrow{D_{\mu}^{(A^{(\rho )}){\dagger}}}=
\bar \psi (x)\loarrow{(\partial_{\mu}-A^{(\rho )}_{\mu}(x))}
\nonumber \\
&&A_{\mu}^{(\rho )}(x)=A_{a\mu}(x)T^a.
\label{4}
\end{eqnarray}

The SU(2) gauge transformations, under which the Lagrangian density is 
invariant, are defined in the following way [$U(x), \tilde U(x), 
U^{(\rho )}(x),$ are their realizations in the adjoint, doublet and $\rho$
representations respectively]

\begin{eqnarray}
&&A_{\mu}(x)\mapsto A^U_{\mu}(x)=U^{-1}(x)A_{\mu}(x)U(x)+U^{-1}(x)\partial_{\mu}
U(x)\nonumber \\
&&F_{\mu\nu}(x)\mapsto F^U_{\mu\nu}(x)=U^{-1}(x)F_{\mu\nu}(x)U(x)\nonumber \\
&&\phi (x)\mapsto \phi^U(x)={\tilde U}^{-1}(x)\phi (x)\nonumber \\
&&\psi (x)\mapsto \psi^U(x)=U^{(\rho ) -1}(x)\psi (x).
\label{5}
\end{eqnarray}

The Euler-Lagrange equations are [``${\buildrel \circ \over =}$" means
evaluated on the extremals of the action $S=\int d^4x {\cal L}(x)$]

\begin{eqnarray}
L_a^{\mu}&=&g^2({{\partial {\cal L}}\over {\partial A_{a\mu}}}-\partial_{\nu}
{{\partial {\cal L}}\over {\partial \partial_{\nu}A_{a\mu}}})={\hat D}^{(A)}
_{\nu ab}F^{\nu\mu}_b+g^2J^{\mu}_a{\buildrel \circ \over =} 0\nonumber \\
&J&^{\mu}_a=i\bar \psi \gamma^{\mu}T^a\psi-
\phi^{\dagger}[{\tilde T}^aD^{(A)\mu}-
\loarrow{D^{(A)\mu {\dagger}}}{\tilde T}^a]\phi \nonumber \\
L_{\psi}&=&{{\partial {\cal L}}\over {\partial \psi}}-\partial_{\mu}
{{\partial {\cal L}}\over {\partial \partial_{\mu}\psi}}=-\bar \psi
[\loarrow{i(\partial_{\mu}-A_{a\mu}T^a)}\gamma^{\mu}+
m]{\buildrel \circ \over =} 0\nonumber \\
L_{\bar \psi}&=&{{\partial {\cal L}}\over {\partial \bar \psi}}-\partial_{\mu}
{{\partial {\cal L}}\over {\partial \partial_{\mu}\bar \psi}}=[\gamma^{\mu}
i(\partial_{\mu}+A_{a\mu}T^a)-m]\psi {\buildrel \circ \over =} 0\nonumber \\
L_{\phi i}&=&{{\partial {\cal L}}\over {\partial \phi_i}}-\partial_{\mu}
{{\partial {\cal L}}\over {\partial \partial_{\mu}\phi_i}}=-
{[D^{(A)\mu}D^{(A)}_{\mu}\phi ]}_i^{\dagger}-{{\partial V(\phi )}\over
{\partial \phi_i}} {\buildrel \circ \over =} 0\nonumber \\
L_{\phi^{*}i}&=&{{\partial {\cal L}}\over {\partial \phi^{*}_i}}-\partial_{\mu}
{{\partial {\cal L}}\over {\partial \partial_{\mu}\phi^{*}_i}}=-
[D^{(A)\mu}D^{(A)}_{\mu}\phi ]_i-{{\partial V(\phi )}\over {\partial 
\phi^{*}_i}}{\buildrel \circ \over =} 0.
\label{6}
\end{eqnarray}

In absence of fermions, the solutions for the gauge potentials and the Higgs 
fields, corresponding to the vanishing of the $\Theta^{oo}(x)$ component of
the energy-momentum tensor [see later on Eq.(\ref{15})] and therefore of the 
total energy, are

\begin{eqnarray}
&&F_{a\mu\nu}(x){\buildrel \circ \over =}0\nonumber \\
&&D^{(A)}_{\mu ij}\phi_j(x){\buildrel \circ \over =} 0\nonumber \\
&&V(\phi ){\buildrel \circ \over =}0\, \Rightarrow \phi^{\dagger}(x)\phi
(x)=\phi_o^2.
\label{7}
\end{eqnarray}

\noindent One such configuration is: $A_{a\mu}(x)=0$ and $\phi (x)=
{\tilde \phi}_o$ [with ${\tilde \phi}_o$ a given doublet, for instance
${\tilde \phi}_o=\left( \begin{array}{l} 0\\ \phi_o \end{array} \right)$].
If in a certain region of spacetime one has $F_{a\mu\nu}(x)\not= \, (or\, =)
0$ but the other two equations (\ref{7}) satisfied, one says  that the fields
are in the ``Higgs vacuum"; see Ref.\cite{go} for the use of this concept in
the theory of non-Abelian monopoles.

Always in absence of fermions, the solution for the Higgs fields which is an
absolute minimum of the potential $V(\phi )$ is (see the next 
Section for the geometrical interpretation)

\begin{eqnarray}
&&D^{(A)}_{\mu ij}\phi_j(x){\buildrel \circ \over =} 0\nonumber \\
&&{{\partial V(\phi )}\over {\partial \phi}} {\buildrel \circ \over =} 0.
\label{8}
\end{eqnarray}

\noindent While the second equation has the two solutions $\phi =0$ and
$\phi =\phi_o$, from the first equation we get $0 {\buildrel \circ \over =}
[D^{(A)}_{\mu},D^{(A)}_{\nu}]\phi (x)=F_{a\mu\nu}(x){\tilde T}^a
\phi (x)$. Therefore, we get, besides the solutions either $\phi =0$ and 
$F\not= 0$ or $\phi =\phi_o$ and $F=0$ of the Abelian case, the solution in 
which the components of $F_{a\mu\nu}$ corresponding to the generators of the
unbroken subgroup $H\subset G$ are non-zero; but in our case H=0. Conditions
like Eqs.(\ref{8}) are imposed on the background fields\cite{sho}, when the
system is studied in an external field ($A_{\mu}\mapsto A^{(ext)}_{\mu}$), 
together with the requirement ${\hat D}^{(A^{(ext)})}_{\mu}F^{(ext)}
_{\nu\rho}(x)=0$ (this implies that the external gauge field is Abelian up to
a gauge transformation).

\section
{Differential Geometric Setting}

From a geometric point of view, see Refs.\cite{fulp,kk}, in spontaneously broken
gauge theories with symmetry-breaking Higgs fields one has: 

i) A principal G-bundle P(M,G) over Minkowski spacetime M [or over one its
fixed $x^o$ slice $R^3$], whose standard fiber is the structure group G
[a compact, semisimple, connected Lie group]. We shall consider only trivial 
principal bundles $P=M\times G$ and simply connected groups G. A connection on 
P is described by a connection one-form $\omega$ over P and with each global
cross section $\sigma :M\rightarrow P=M\times G$ is associated a global gauge
potential $A_{\mu}(x)dx^{\mu}=A_{a\mu}(x){\hat T}^adx^{\mu}=\sigma^{*}\omega$,
which is a one-form over M. The group ${\cal G}$ of gauge transformations
connects all the gauge potentials on the same gauge orbit, associated with the 
given connection one-form $\omega$ on P, by considering all the possible
global cross sections $\sigma :M\rightarrow P$ (therefore $P=M\times G$ can 
also be considered as the group manifold of ${\cal G}$).

ii) A  bundle $E_{\psi}$, associated to $P=M\times G$, over M with the same
structure group G and whose standard fiber is the vector space (with
Grassmann-valued vectors as elements) of a representation $\rho$ of G with
generators $T^a$; its global cross sections describe the fermion fields $\psi 
(x)$ of the model.

iii) A bundle E, associated to $P=M\times G$, over M with the same structure
group G, whose standard fiber F is the vector space of a representation
$\rho^{'}$ of G with generators ${\tilde T}^a$; its global cross sections are
the complex scalar (generalized) Higgs fields on P of the model\cite{tra}, which
can also be described by special maps $\tilde \phi :P\rightarrow F$ [called
tensorial 0-forms in Ref.\cite{kn}]; the usual Higgs fields on M are
$\phi :M\rightarrow F$, $\phi =\sigma^{*}\tilde \phi$, $\sigma :M\rightarrow P$.
When F is only a manifold (usually it is a vector space) with a G-action, the
(generalized) Higgs field $\tilde \phi$ is called a ``symmetry-breaking Higgs 
field" if $\tilde \phi :P\rightarrow F$ maps all of P onto a single G-orbit 
$\xi$ with $dim\, \xi \geq 1$ (more in general onto a union of G-orbits) of the
G-action of G on F and each point of the G-orbit $\xi$ has an isotropy
(or little or stability) group $H \subset G$ under the G-action; the 
spontaneously broken
symmetries are the elements $a\in G$, $a\notin H$, because
$a: \phi_o\in \xi \mapsto \phi \not= \phi_o$, where $\phi_o$ is a reference 
point in $\xi$ (called a ``vacuum state" at the quantum level). This means
that $\tilde \phi:P\rightarrow F$ identifies in $P=M\times G$ a residual 
symmetry group $H\subset G$ [$b\in H \Rightarrow b: \phi_o\in \xi \mapsto
\phi_o$] and a corresponding subbundle $Q=M\times H$ [Q 
must be such to satisfy the fundamental hypothesis that $g=h\oplus {\cal M}$ 
with $g$ the Lie algebra of G, $h$ the Lie algebra of H and ${\cal M}$ a 
complementary space (with respect to an adjoint-invariant scalar product on
$g$) invariant under the adjoint action of H].

On M one adds to the Yang-Mills plus fermion Lagrangian density a term
${[D^{(A)}_{\mu}\phi ]}^{\dagger}D^{(A)\mu}\phi +V(\phi )$ with $V(\phi )$ a
suitable potential and looks for the subset of solutions of the Higgs 
Euler-Lagrange  equations satisfying $D^{(A)}_{\mu}\phi {\buildrel \circ 
\over =} 0$ and ${{\partial V(\phi )}\over {\partial \phi}}{\buildrel \circ 
\over =} 0$; the potential $V(\phi )$ must be such that it admits an
absolute mininum (assumed equal to zero) for all those values of $\phi$ 
corresponding to symmetry-breaking Higgs fields: namely $V^{-1}(0)$ is spanned 
by $\phi =\sigma^{*}\tilde \phi$ with $\tilde \phi :P\rightarrow \xi \subset 
F$. The condition $D^{(A)}_{\mu}\phi =0$ \cite{tra} of being covariantly 
constant with respect to the connection $\omega$ associated with the gauge 
potential $A_{\mu}$ is the necessary and sufficient condition\cite{kn} so
that the connection $\omega$ on $P=M\times G$ be reducible to a connection 
$\omega_o$ on the subbundle $Q=M\times H$. If $i:Q\rightarrow P$ is the 
inclusion map, one has $i^{*}\omega =\omega_o+\gamma$ with $\omega_o$ a
connection one-form on Q and $\gamma$ being called a ``tensorial one-form of
type $\mu$", where $\mu$ is the representation of H on ${\cal M}$ induced by the
representation $\rho^{'}$ of G. See Ref.\cite{isham}, for the description
of spontaneous symmetry breaking with the Higgs mechanism in the context of
general relativity.

In physical terms\cite{cl,or} $\gamma$ describes the massive gauge fields (whose
mass depends on the representation $\rho^{'}$ of G), while $\omega_o$ is the
connection associated with those gauge potentials which remain massless; if
$A_{a\mu}(x)$ are the original gauge potentials, with $a=1,..,dim\, g$, and if
$b=1,..,dim\, h$ labels the generators of $h$, $\lbrace A_{1\mu},.., A_{dim\,
h\, \mu}\rbrace$ are the massless gauge potentials and $\lbrace A_{dim\, h+1\, 
\mu},..,A_{dim\, g\, \mu}\rbrace$ are the massive ones.

Note that in this geometrical construction one never speaks of massless
Goldstone bosons (with the quantum numbers of the broken generators at the
quantum level) associated with the spontaneous symmetry breaking from G to H
(here G is regarded as the global rigid symmetry group contained in the
group of gauge transformations ${\cal G}$\cite{lus}); as shown in 
Ref.\cite{str} the Goldstone bosons (and the associated infrared singularities) 
are hidden in the unphysical gauge degrees of freedom of the model (they are a
subset of the Higgs fields) present in gauge theories due to the Gauss'
laws. As we shall see in the last Section, the discussion of the Gauss law 
first class constraints is not trivial as in absence of spontaneous symmetry 
breaking with the Higgs mechanism: the reduction to Dirac's observables
(equivalent to the unitary gauge but without the introduction of any 
gauge-fixing) elucidates the real meaning of the statement ``the would-be
Goldstone bosons are eaten by those gauge bosons which become massive". After
the reduction one is left with only a subset of physical Higgs fields which
depend on the original representation $\rho^{'}$ of G.

In this paper, we shall consider the simplest case of G=SU(2) and H=0, so
that $P=M\times G$ is reduced to $Q=M\times \lbrace 0\rbrace$ and no massles
gauge fields are left.

\section
{The Hamiltonian Formalism}

The canonical momenta implied by the Lagrangian density (\ref{1}) are

\begin{eqnarray}
&&\pi^o_a(x)={{\partial {\cal L}(x)}\over {\partial \partial_oA_{ao}(x)}}=0
\nonumber \\
&&\pi^k_a(x)={{\partial {\cal L}(x)}\over {\partial \partial_oA_{ak}(x)}}=
-g^{-2}F_a^{ok}(x)=g^{-2}E^k_a(x)\nonumber \\
&&\pi_{A\alpha}(x)={{\partial {\cal L}(x)}\over {\partial \partial_o
\psi_{A\alpha}(x)}}=-{i\over 2}{(\bar \psi (x)\gamma_o)}_{A\alpha}\nonumber \\
&&{\bar \pi}_{A\alpha}(x)={{\partial {\cal L}(x)}\over {\partial \partial_o
{\bar \psi}_{A\alpha}(x)}}=-{i\over 2}{(\gamma_o\psi (x))}_{A\alpha}\nonumber \\
&&\pi_{\phi \, i}(x)={{\partial {\cal L}(x)}\over {\partial \partial_o
\phi_i(x)}}={[D^{(A)}_o\phi (x)]}_i^{\dagger}\nonumber \\
&&\pi_{\phi^{\dagger}\, i}(x)={{\partial {\cal L}(x)}\over {\partial \partial_o
\phi_i^{*}(x)}}={[D^{(A)}_o\phi (x)]}_i.
\label{9}
\end{eqnarray}

They satisfy the standard Poisson brackets

\begin{eqnarray}
&&\lbrace A_{a\mu}(\vec x,x^o),\pi_b^{\nu}(\vec y,x^o)\rbrace =\delta_{ab}
\delta^{\nu}_{\mu}\delta^3(\vec x-\vec y)\nonumber \\
&&\lbrace \psi_{A\alpha}(\vec x,x^o),\pi_{B\beta}(\vec y,x^o)\rbrace =
\lbrace {\bar \psi}_{A\alpha}(\vec x,x^o),{\bar \pi}_{B\beta}(\vec y,x^o)
\rbrace =-\delta_{AB}\delta_{\alpha\beta}\delta^3(\vec x-\vec y)\nonumber \\
&&\lbrace \phi_i(\vec x,x^o),\pi_{\phi \, j}(\vec y,x^o)\rbrace =
\lbrace \phi^{*}_i(\vec x,x^o),\pi_{\phi^{\dagger}\, j}(\vec y,x^o)\rbrace =
\delta_{ij}\delta^3(\vec x-\vec y).
\label{10}
\end{eqnarray}

Following Ref.\cite{lus}, the second-class constraints $\pi_{A\alpha}(x)+{i
\over 2}{(\bar \psi (x)\gamma_o)}_{A\alpha}\approx 0$, ${\bar \pi}_{A\alpha}(x)+
{i\over 2}{(\gamma_o\psi (x))}_{A\alpha}\approx 0$, are eliminated by going to
Dirac brackets; then the surviving variables $\psi_{A\alpha}(x), {\bar \psi}
_{A\alpha}(x)$ satisfy (for the sake of simplicity we still use the notation
$\lbrace .,.\rbrace$ for the Dirac brackets)

\begin{eqnarray}
&&\lbrace \psi_{A\alpha}(\vec x,x^o),{\bar \psi}_{B\beta}(\vec y,x^o)\rbrace
=-i{(\gamma^o)}_{\alpha\beta}\delta_{AB}\delta^3(\vec x-\vec y)\nonumber \\
&&\lbrace \psi_{A\alpha}(\vec x,x^o),\psi_{B\beta}(\vec y,x^o)\rbrace =
\lbrace {\bar \psi}_{A\alpha}(\vec x,x^o),{\bar \psi}_{B\beta}(\vec y,x^o)
\rbrace =0.
\label{11}
\end{eqnarray}

The resulting Dirac Hamiltonian density is (after an allowed integration by 
parts; $\lambda_{ao}(x)$ is a Dirac multiplier)

\begin{eqnarray}
{\cal H}_D(x)&=&{1\over 2}\sum_a[g^2{\vec \pi}_a^2(x)+g^{-2}{\vec B}_a^2(x)]+
\nonumber \\
&+&{i\over 2}\psi^{\dagger}(x)[\vec \alpha \cdot (\vec \partial +{\vec A}^{(\rho
)}(x))-\loarrow{(\vec \partial -{\vec A}^{(\rho )}(x))}\cdot
\vec \alpha ]\psi (x) 
+m\bar \psi (x)\psi (x)+\nonumber \\
&+&\pi_{\phi {*}i}(x)\pi_{\phi i}(x)+[{\vec D}^{(A){\dagger}}\phi^{*}(x)]_i
\cdot [{\vec D}^{(A)}\phi (x)]_i
+\lambda {(\phi^{\dagger}(x)\phi (x)-\phi_o^2)}^2-\nonumber \\
&-&A_{ao}(x)[-\vec \partial \cdot {\vec \pi}_a(x)-c_{abc}{\vec A}_b(x)\cdot 
{\vec \pi}_c(x)+i\psi^{\dagger}(x)T^a\psi (x)+\nonumber \\
&+&{i\over 2}(\pi_{\phi i}(x)(\tau^a)_{ij}\phi_j(x)-\pi_{\phi {*}i}(x)
(\tau^a)_{ij}\phi_j^{*}(x))]+\lambda_{ao}(x)\pi^o_a(x).
\label{12}
\end{eqnarray}

\noindent where $B^k_a=-{1\over 2}\epsilon^{ijk}F^{ij}_a$.

The time constancy of the primary constraints $\pi^o_a(x)\approx 0$ yields
the Gauss law secondary constraints

\begin{eqnarray}
\Gamma_a(x)&=&-\vec \partial \cdot {\vec \pi}_a(x)-c_{abc}{\vec A}_b(x)\cdot
{\vec \pi}_c(x)+i\psi^{\dagger}(x)T^a\psi (x)+\nonumber \\
&+&{i\over 2}[\pi_{\phi \, i}(x){(\tau^a)}_{ij}\phi_j(x)-\pi_{\phi^{*}\, j}
{(\tau^a)}_{ij}\phi^{*}_j(x)]\approx 0.
\label{13}
\end{eqnarray}

The $\Gamma_a(x)$'s are constants of the motion and the six constraints 
$\pi^o_a(x)\approx 0$, $\Gamma_a(x)\approx 0$ are first class with the only 
nonvanishing Poisson brackets

\begin{equation}
\lbrace \Gamma_a(\vec x,x^o),\Gamma_b(\vec y,x^o)\rbrace =c_{abc}\Gamma_c
(\vec x,x^o) \delta^3(\vec x-\vec y)
\label{14}
\end{equation}

The equations $\Gamma_a(x)=0$, namely the acceleration independent 
Euler-Lagrange equations of the model, are ambiguous. They can be thought either
as elliptic equations in the non-Abelian electric fields ${\vec \pi}_a(x)=
g^{-2}{\vec E}_a(x)$ or as algebraic equations for three of the Higgs
momenta $\pi_{\phi \, i}(x), \pi_{\phi^{*} \, i}(x)$. Since each equation has
this ambiguity, we find that the model admits eight (modulo
identifications)  disjoint sets of solutions
(the space of solutions has a nontrivial zeroth homotopy group) of the
Gauss laws and, therefore, eight inequivalent phases. Only two of these phases
preserve SU(2) covariance: i) the SU(2) phase, in which all the three
equations are solved in ${\vec \pi}_a(x)$ and in which the SU(2) fields remain
massless; ii) The Higgs phase, in which all the equations are solved in the 
Higgs momenta and the SU(2) fields become massive (spontaneous symmetry 
breaking through the Higgs mechanism). See after Eq.(\ref{24}) for more
details.

Let us make a digression on the choice of the boundary conditions on the
various fields.
Since the conserved energy-momentum and angular momentum tensor densities 
and Poincar\'e generators are 
[$\sigma^{\mu\nu}={i\over 2}[\gamma^{\mu},\gamma^{\nu}]$, $\sigma^i={1\over 2}
\epsilon^{ijk}\sigma^{jk}$, $\vec \alpha =\gamma^o\vec \gamma$, $\beta =
\gamma^o$]

\begin{eqnarray}
\Theta^{\mu\nu}(x)&=&g^{-2}(F_a^{\mu\alpha}(x)F_{a\alpha}{}^{\nu}(x)+{1\over 4}
\eta^{\mu\nu}F_a^{\alpha\beta}(x)F_{a\alpha\beta}(x))+\nonumber \\
&+&{i\over 2}\bar \psi (x)[\gamma^{\mu}D^{(A^{(\rho )})\nu}-
\loarrow{D^{(A^{(\rho )}){\dagger}\nu}}\gamma^{\mu}]\psi (x)+
\nonumber \\
&+&{(D^{(A)\mu}\phi (x))}^{\dagger}\, D^{(A)\nu}\phi (x)+{(D^{(A)\nu}\phi (x))}
^{\dagger}D^{(A)\mu}\phi (x)-\nonumber \\
&-&\eta^{\mu\nu}[{(D^{(A)\alpha}\phi (x))}^{\dagger}D^{(A)}{}_{\alpha}
\phi (x)-V(\phi )],\nonumber \\
{\cal M}^{\mu\alpha\beta}(x)&=&x^{\alpha}\Theta^{\mu\beta}(x)-x^{\beta}
\Theta^{\mu \alpha}(x)+{1\over 4}\bar \psi (x)(\gamma^{\mu}\sigma^{\alpha\beta}
+\sigma^{\alpha\beta}\gamma^{\mu})\psi (x),\nonumber \\
&&\partial_{\nu}\Theta^{\nu\mu}(x){\buildrel \circ \over =} 0,\quad\quad
\partial_{\mu}{\cal M}^{\mu\alpha\beta}(x){\buildrel \circ \over =} 0,
\nonumber \\
&&{}\nonumber \\
P^{\mu}&=&\int d^3x \Theta^{o\mu}(\vec x,x^o),\nonumber \\
J^{\mu\nu}&=&\int d^3x {\cal M}^{o\mu\nu}(\vec x,x^o),\nonumber \\
&&{}\nonumber \\
P^o&=&\int d^3x\, \lbrace {1\over 2}\sum_a[g^2{\vec \pi}_a^2(\vec x,x^o)+
g^{-2}{\vec B}_a^2(\vec x,x^o)]+\nonumber \\
&+&\pi_{\phi}(\vec x,x^o)\pi_{\phi^{*}}(\vec x,x^o)+{({\vec D}^{(A)}\phi
(\vec x,x^o))}^{\dagger}\cdot {\vec D}^{(A)}\phi (\vec x,x^o)+V(\phi )
\nonumber \\
&+&{i\over 2}\psi^{\dagger}(\vec x,x^o)[D^{(A^{(\rho )})o}
-\loarrow{D^{(A^{(\rho )}){\dagger}o}}]\psi (\vec x,x^o)\rbrace \nonumber \\
P^i&=&\int d^3x\, \lbrace {({\vec \pi}_a (\vec x,x^o)\times {\vec B}_a(\vec x,
x^o))}^i+\nonumber \\
&+&\pi_{\phi}(\vec x,x^o)D^{(A)i}\phi (\vec x,x^o)+{(D^{(A)i}\phi
(\vec x,x^o))}^{\dagger}\pi_{\phi^{*}}(\vec x,x^o)+\nonumber \\
&+&{i\over 2}\psi^{\dagger}(\vec x,x^o)[D^{(A^{(\rho )})i}
+\loarrow{D^{(A^{(\rho )}{\dagger}i}}]\psi (\vec x,x^o)\rbrace \nonumber \\
J^i&=&{1\over 2}\epsilon^{ijk}J^{jk}=\int d^3x\, \lbrace {[\vec x\times({\vec
\pi}_a(\vec x,x^o)\times {\vec B}_a (\vec x,x^o))]}^i+\nonumber \\
&+&{[\vec x\times (\pi_{\phi}(\vec x,x^o){\vec D}^{(A)}\phi (\vec x,x^o)+
({\vec D}^{(A)}\phi (\vec x,x^o))^{\dagger}\pi_{\phi^{*}}(\vec x,x^o))]}^i+
\nonumber \\
&+&{i\over 2}\psi^{\dagger}(\vec x,x^o)[\vec x\times ({\vec D}^{(A^{(\rho )})}
+\loarrow{{\vec D}^{(A^{(\rho )}{\dagger}}})]^i\psi (\vec x,x^o)+\nonumber \\
&+&{1\over 2}\psi^{\dagger} (\vec x,x^o)\sigma^i
\psi (\vec x,x^o)\rbrace \nonumber \\
K^i&=&J^{oi}=x^oP^i-\int d^3x\, x^i\Theta^{oo}(\vec x,x^o),
\label{15}
\end{eqnarray}

\noindent
following Ref.\cite{lus}, we will assume boundary conditions 
[$r=\, |\, \vec x\, |$] $A_{ao}(\vec x,x^o){\rightarrow}_{r\rightarrow
\infty}\, a_o/r^{1+\epsilon}$, ${\vec A}_a(\vec x,x^o){\rightarrow}
_{r\rightarrow \infty}{\vec a}_a/r^{2+\epsilon}$,  
$\pi^o_a(\vec x,x^o){\rightarrow}_{r\rightarrow \infty}\, p^o_a/r^{1+\epsilon}
+O(r^{-2})$, ${\vec \pi}_a(\vec x,x^o){\rightarrow}_{r\rightarrow \infty}\,
{\vec e}_a/r^{2+\epsilon}+O(r^{-3})$, $\lambda_{ao}(\vec x,x^o)
{\rightarrow}_{r\rightarrow \infty}\, e_{ao}/r^{1+\epsilon}+O(r^{-2})$,
$\psi(\vec x,x^o){\rightarrow}_{r\rightarrow \infty}\, \chi/r^{3/2+\epsilon}+
O(r^{-2})$, $\phi (\vec x,x^o){\rightarrow}_{r\rightarrow \infty}\, const.+
\varphi /r^{2+\epsilon}+O(r^{-3})$ [the constant allows the existence of minima
$\phi_o$ of the potential $V(\phi )$],  $\pi_{\phi}(\vec x,x^o){\rightarrow}
_{r\rightarrow  \infty}\, \zeta /r^{2+\epsilon}+O(r^{-3})$,
so that $\Gamma_a(\vec x,x^o){\rightarrow}_{r\rightarrow \infty}\, \gamma_a
/r^{3+\epsilon}+O(r^{-4})$ and the Poincar\'e generators are finite.
If we assume that the gauge transformations behave as $U(\vec x,x^o)
{\rightarrow}_{r\rightarrow \infty}\, const. + O(r^{-1})$, $U^{(\rho )}(\vec x,
x^o){\rightarrow}_{r\rightarrow \infty}\, const.\, +O(r^{-1})$, $\tilde U
(\vec x,x^o){\rightarrow}_{r\rightarrow \infty}\, const. \openone +O(r^{-1})$,
the previous boundary
conditions on the fields are preserved by the gauge transformations and the
non-Abelian SU(2) charges (see the last Section) transform covariantly
under them\cite{lus}. Let us remark that the previous boundary conditions
are adapted to the fixed $x^o$, not Lorentz covariant, Hamiltonian formalism;
however, they become natural in its covariantization based on the
reformulation of the theory on spacelike hypersurfaces\cite{lus,lus2,dir,lv}.

Let us make a technical remark about the choice of the function space the
Yang-Mills gauge potentials belong to in spontaneously broken gauge theories
with the Higgs mechanism. As in Ref.\cite{lus}, we consider only trivial 
principal bundles and we exclude monopole solutions, but now it is not clear
what to do with the Gribov ambiguity, because the condition $D^{(A)}_{\mu}
\phi (x)=0$ of Eq.(\ref{8}) for the reducibility of the connection $\omega$
on $P=M\times G$ to a connection $\omega_o$ on the subbundle $Q=M\times H$
(here $Q=M\times \lbrace 0\rbrace$) implies the existence of gauge symmetries
(nontrivial stability group of a gauge potential) so that the needed space of
connections on $P=M\times G$  cannot contain only fully irreducible connections
[as shown in Ref.\cite{lus}, only in this case the Gribov ambiguity is absent
(the stability groups of the gauge potentials (gauge symmetries) and of the
field strengths (gauge copies) are trivial) and this is obtainable by the 
choice of special weighthed Sobolev spaces]. If we formulate the theory in
ordinary Sobolev spaces, as it is usually done, we do not have problems in the 
Higgs phase, because there the reduction associated with the Gauss' laws first
class constraints is purely algebraic. The problem with the Gribov ambiguity
arises in the reduction of the SU(2) phase and of the mixed non-SU(2)-covariant
ones. However these phases are not physical, so that we do not worry about them
[however there can be a cosmological use \cite{cosmo}, for explaining the
observed cosmological baryon density, of the phase transition
restoring the ordered SU(2) phase from the disordered Higgs one (what about
the mixed phases?)]; see also Ref.\cite{sho} for the phase transition
restoring the SU(2) symmetry in presence of a constant external electromagnetic
field. The only problem is that the formal proofs of
renormalizability need all the phases and, therefore, they have to face the
problems connected with the Gribov ambiguity.

\section
{The Higgs phase}

In this paper we shall study only the Higgs phase, because the SU(2) phase
can be reduced by combining the methods of Ref.\cite{lv,lus}.

The parametrization of the Higgs fields suitable to the Higgs phase and 
realizing the spontaneous symmetry breaking by a choice of a reference
point in the degenerate set of minima of the classical
potential $V(\phi )$ [$\phi^{\dagger}\phi =\phi_o^2$] is

\begin{eqnarray}
\phi (x)&=&e^{{\tilde T}^a\theta_a(x)}\, \left( \begin{array}{l} 0\\
\phi_o+{1\over {\sqrt{2}}}H(x) \end{array} \right)=e^{{\tilde T}^a\theta_a(x)}\,
{1\over {\sqrt{2}}} \left( \begin{array}{l} 0\\ v+H(x)\end{array} \right)=
{\tilde U}_{\theta}(x)\tilde \phi (x)\nonumber \\
&&{\tilde T}^a\theta_a(x)=-i{{\vec \tau}\over 2}\cdot \vec \theta (x),\quad\quad
v=\sqrt{2}\phi_o,\quad\quad \tilde \phi (x)=[\phi_o+{1\over {\sqrt{2}}}H(x)]
\left( \begin{array}{ll} 0\\ 1\end{array} \right),
\label{16}
\end{eqnarray}

\noindent with $\theta_a(x), H(x)$ real fields [$H(\vec x,x^o)
{\rightarrow}_{r\rightarrow \infty}\, h/r^{2+\epsilon}+O(r^{-3})$, $\theta_a
(\vec x,x^o){\rightarrow}_{r\rightarrow \infty}\, \zeta_a/r^{2+\epsilon}+
O(r^{-3})$].

The value $\phi =0$ is not covered by these radial coordinates; for the sake 
of simplicity we take a positive value $\phi_o > 0$ for the arbitrary symmetry 
breaking reference point in the set of minima of the potential: this set is 
spanned by varying the angular variables $\theta_a$, so that the $\theta_a$'s 
are the would-be Goldstone bosons; the symmetry group SU(2) is completely
broken and there is no residual stability group of the points of minimum.

The parametrization of Eq.(\ref{16}) requires a restriction to Higgs
fields which have no zeroes, namely $\phi^{\dagger}(x)\phi (x)\not= 0$ 
[$H(x)\not= -v=-\sqrt{2}\phi_o$], and with nonsingular phases $\theta_a(x)$'s 
because we assumed a trivial SU(2) principal bundle. The analogue of the 
quantum statement of symmetry breaking, i.e. that the theory is invariant 
under a group G but not the ground state, is replaced by the choice of the 
parametrization (\ref{16}) with a given $\phi_o$, i.e. by the choice of a 
family of solutions of the Euler-Lagrange equations associated with 
Eq.(\ref{1}) not invariant under SU(2).

Since the parametrization has the form of a gauge transformation, we have

\begin{eqnarray}
D^{(A)}_{\mu}\phi (x)&=&{\tilde U}_{\theta}(x)D_{\mu}^{(A^{U_{\theta}})}\tilde
\phi (x)={\tilde U}_{\theta}(x) [{1\over {\sqrt{2}}}\partial_{\mu}H(x)+
(\phi_o+{1\over {\sqrt{2}}}H(x)){\tilde A}
_{\mu}^{U_{\theta}}(x)] \left( \begin{array}{l} 0\\ 1\end{array} \right) ,
\nonumber \\
{[D_{\mu}^{(A)}\phi (x)]}^{\dagger}&=&\left( \begin{array}{ll} 0& 1
\end{array} \right) [{1\over {\sqrt{2}}}\partial_{\mu}H(x)-(\phi_o+
{1\over {\sqrt{2}}}H(x)){\tilde A}_{\mu}^{U_{\theta}}(x)]
{\tilde U}^{-1}_{\theta}(x),\nonumber \\
&&\phi^{\dagger}\phi ={\tilde \phi}^{\dagger}\tilde \phi =(\phi_o+{1\over
{\sqrt{2}}}H)^2.
\label{17}
\end{eqnarray}

By using $\tau^a\tau^b=\delta_{ab}+i\epsilon_{abc}\tau^c$, 
$\left( \begin{array}{ll} 0& 1\end{array} \right) {\tilde T}^c
{\tilde T}^d\left( \begin{array}{l}0\\ 1\end{array} \right) =-{1\over 4}(1+
\tau^2)_{cd}$, $e^{\theta_b{\tilde T}^b}=e^{-{i\over 2}\theta {\hat n}_b\tau^b}
= cos\, {{\theta}\over 2} -i sin\, {{\theta}\over 2} {\hat n}_b \tau^b$
[with $\sum_b {\hat n}^2_b =1$, $\theta = \sqrt{\theta^2_1+\theta^2_2+
\theta_3^2}$, ${\hat n}_a=\theta_a/\theta$], we get

\begin{eqnarray}
{\tilde A}^{U_{\theta}}_{\mu}(x)&=&{\tilde U}^{-1}_{\theta}(x) [ {\tilde A}
_{a\mu}(x){\tilde T}^a+\partial_{\mu}] {\tilde U}_{\theta}(x)=\nonumber \\
&=&\lbrace [ cos\, \theta (x) \delta_{ab} + 2 sin^2\, {{\theta (x)}\over 2}
{\hat n}_a(x) {\hat n}_b(x) +sin\, \theta (x) \epsilon_{abc}{\hat n}_c(x)]
A_{b\mu}(x)+\nonumber \\
&+& [{\hat n}_a(x) {{\partial \theta (x)}\over {\partial \theta_b}} + sin\,
\theta (x) {{\partial {\hat n}_a(x)}\over {\partial \theta_b}}-2 sin^2\,
{{\theta (x)}\over 2} \epsilon_{acd}{\hat n}_c(x) {{\partial {\hat n}_d(x)}\over
{\partial \theta_b}}] \partial_{\mu}\theta_b(x) \rbrace {\tilde T}^a,
\label{18}
\end{eqnarray}

\begin{eqnarray}
&&\left( \begin{array}{ll} 0& 1\end{array} \right)
[{\tilde A}^{U_{\theta}}_{\mu}(x){\tilde A}^{U_{\theta}\, \mu}(x)]
\left( \begin{array}{l}0\\ 1\end{array} \right)=\nonumber \\
&=&-{1\over 4} [A_{a\mu}(x)+\nonumber \\
&+&({\hat n}_a(x) {{\partial \theta (x)}\over 
{\partial \theta_b}}+sin\, \theta (x) {{\partial {\hat n}_a(x)}\over {\partial
\theta_b}} +2 sin^2\, {{\theta (x)}\over 2} \epsilon_{acd}{\hat n}_c(x)
{{\partial {\hat n}_d(x)}\over {\partial \theta_b}}) \partial_{\mu} \theta_b
(x) ]^2,
\label{19}
\end{eqnarray}

\noindent so that the Lagrangian density (\ref{1}) becomes

\begin{eqnarray}
{\cal L}(x)&=&-{1\over {4g^2}}F_{a\mu\nu}(x)F_a^{\mu\nu}(x)+\nonumber \\
&+&{1\over 4} (\phi_o+{1\over {\sqrt{2}}}H(x))^2\times \nonumber \\
&&[A_{a\mu}(x)+({\hat n}_a(x) {{\partial \theta (x)}\over 
{\partial \theta_b}}+sin\, \theta (x) {{\partial {\hat n}_a(x)}\over {\partial
\theta_b}} +2 sin^2\, {{\theta (x)}\over 2} \epsilon_{acd}{\hat n}_c(x)
{{\partial {\hat n}_d(x)}\over {\partial \theta_b}}) \partial_{\mu} \theta_b
(x) ]^2+\nonumber \\
&+&{1\over 2}\partial_{\mu}H(x)\partial^{\mu}H(x)-
{{\lambda}\over 2} H^2(x)({1\over \sqrt{2}}H(x)+2\phi_o)^2+\nonumber \\
&+&{i\over 2}\bar \psi (x)[\gamma^{\mu}(\partial_{\mu}+A_{a\mu}(x)T^a)-
\loarrow{(\partial_{\mu}-A_{a\mu}(x)T^a)}\gamma^{\mu}]
\psi (x)-m\bar \psi (x)\psi (x),
\label{20}
\end{eqnarray}

The new Higgs canonical momenta are

\begin{eqnarray}
\pi_H(x)&=&{{\partial {\cal L}(x)}\over {\partial \partial_oH(x)}}=
\partial^oH(x)\nonumber \\
\pi_{\theta_a}(x)&=&{{\partial {\cal L}(x)}\over {\partial \partial_o\theta_a
(x)}}={1\over 2}[\phi_o+{1\over {\sqrt{2}}}H(x)]^2 \times \nonumber \\
&&V_{ab}(\theta (x)) [{\tilde A}_{ob}(x) +V_{bc}(\theta (x)) \partial_o
\theta_c(x)],\nonumber \\
&&{}\nonumber \\
V_{ab}(\theta (x))&=&{\hat n}_a(x) {{\partial \theta (x)}\over {\partial 
\theta_b}} +sin\, \theta (x) {{\partial {\hat n}_a(x)}\over {\partial \theta_b}}
+2sin^2\, {{\theta (x)}\over 2} \epsilon_{acd}{\hat n}_c(x) {{\partial {\hat
n}_d(x)}\over {\partial \theta_b}} =\nonumber \\
&=&{\hat n}_a(x){\hat n}_b(x)+{{sin\, \theta (x)}\over {\theta (x)}} (\delta
_{ab}-{\hat n}_a(x){\hat n}_b(x)) -2 sin^2\, {{\theta (x)}\over 2} \epsilon
_{abc} {{{\hat n}_c(x)}\over {\theta (x)}},\nonumber \\
V^{-1}_{ab}(\theta (x))&=&{\hat n}_a(x){\hat n}_b(x)+{{\theta (x)}\over 2}
cot\, {{\theta (x)}\over 2} (\delta_{ab}-{\hat n}_a(x){\hat n}_b(x))+
{{\theta (x)}\over 2} \epsilon_{abc}{\hat n}_c(x)\nonumber \\
&&{}\nonumber \\
&\Rightarrow& \partial_o\theta_a(x) =V^{-1}_{ab}(\theta (x)) [ V^{-1}_{bc}
(\theta (x)) {{2\pi_{\theta_c}(x)}\over {(\phi_o+{1\over {\sqrt{2}}}H(x))^2}}
-{\tilde A}_{ob}(x)].
\label{21}
\end{eqnarray}

The relation between the old and the new Higgs momenta is

\begin{eqnarray}
\pi_{\phi}(x)&=&
\left( \begin{array}{ll} 0& 1\end{array} \right) [{1\over {\sqrt{2}}}\pi
_H(x)-(\phi_o+{1\over {\sqrt{2}}}H(x))\times \nonumber \\
&&[cos\, \theta (x) \delta_{ab}+2sin^2\, {{\theta (x)}\over 2} {\hat n}_a(x)
{\hat n}_b(x) +sin\, \theta (x) \epsilon_{abc}{\hat n}_c(x)]{\tilde A}_{bo}(x)
+\nonumber \\
&+&[{\hat n}_a(x){{\partial \theta (x)}\over {\partial \theta_b}}+sin\, \theta 
(x) {{\partial {\hat n}_a(x)}\over {\partial \theta_b}} -2sin^2\, {{\theta
(x)}\over 2} \epsilon_{acd}{\hat n}_c(x) {{\partial {\hat n}_d(x)}\over
{\partial \theta_b}}] V^{-1}_{br}(\theta (x)) \times \nonumber \\
&&[V^{-1}_{rs}(\theta (x)) {{2\pi_{\theta_s}(x)}\over {(\phi_o+{1\over
{\sqrt{2}}}H(x))^2}} -{\tilde A}_{ro}(x)] \rbrace {\tilde T}^a]
e^{-\theta_d(x){\tilde T}^d}\nonumber \\
\pi_{\phi^{*}}(x)&=&e^{\theta_d(x){\tilde T}^d}
[{1\over {\sqrt{2}}}\pi
_H(x)+(\phi_o+{1\over {\sqrt{2}}}H(x))\times \nonumber \\
&&[cos\, \theta (x) \delta_{ab}+2sin^2\, {{\theta (x)}\over 2} {\hat n}_a(x)
{\hat n}_b(x) +sin\, \theta (x) \epsilon_{abc}{\hat n}_c(x)]{\tilde A}_{bo}(x)
+\nonumber \\
&+&[{\hat n}_a(x){{\partial \theta (x)}\over {\partial \theta_b}}+sin\, \theta 
(x) {{\partial {\hat n}_a(x)}\over {\partial \theta_b}} -2sin^2\, {{\theta
(x)}\over 2} \epsilon_{acd}{\hat n}_c(x) {{\partial {\hat n}_d(x)}\over
{\partial \theta_b}}] V^{-1}_{br}(\theta (x)) \times \nonumber \\
&&[V^{-1}_{rs}(\theta (x)) {{2\pi_{\theta_s}(x)}\over {(\phi_o+{1\over
{\sqrt{2}}}H(x))^2}} -{\tilde A}_{ro}(x)] \rbrace {\tilde T}^a]
\left( \begin{array}{l}0\\ 1\end{array} \right) .
\label{22}
\end{eqnarray}

The new Dirac Hamiltonian density is

\begin{eqnarray}
{\cal H}_D(x)
&=& {1\over 2}\sum_a[g^2{\vec \pi}^2_a(x)+g^{-2}{\vec B}_a^2(x)]+\nonumber \\
&+&[2V^{-1}_{ac}(\theta (x))-V^{-1}_{ca}(\theta (x))] V^{-1}_{cb}(\theta (x))
{{\pi_{\theta_a}(x) \pi_{\theta_b}(x)}\over {(\phi_o+{1\over {\sqrt{2}}}H(x))
^2}}+\nonumber \\
&+&{1\over 4}[\phi_o+{1\over {\sqrt{2}}}H(x)]^2
\sum_a[{\vec {\tilde A}}_a(x)+V_{ab}(\theta (x)) \vec \partial \theta_b
(x)]^2+\nonumber \\
&+&{i\over 2}\bar \psi (x)[\vec \gamma \cdot (\vec \partial +{\vec A}_a(x)T^a)-
\loarrow{(\vec \partial -{\vec A}_a(x)T^a)}\cdot \vec \gamma ]\psi (x)+
m\bar \psi (x)\psi (x)-\nonumber \\
&-&A_{ao}(x)[-{\hat {\vec D}}^{(A)}_{ab}\cdot {\vec \pi}_b(x)+i\bar \psi (x)
\gamma^oT^a\psi (x)+\pi_{\theta_b}(x)V^{-1}_{ba}(\theta (x))]+
\lambda_{ao}(x)\pi^o_a(x).
\label{23}
\end{eqnarray}

The Gauss law secondary constraints take the form

\begin{equation}
\Gamma_a(x)=-\vec \partial \cdot {\vec \pi}_a(x)-c_{abc}{\vec A}_b(x)\cdot
{\vec \pi}_c(x)+i\psi^{\dagger}(x)T^a\psi (x)+\pi_{\theta_b}(x)V^{-1}_{ba}
(\theta (x))\approx 0
\label{24}
\end{equation}

\noindent which can be trivially solved in the Higgs momenta $\pi_{\theta_a}
(x)$.

We can now see explicitely the existing phases:

\noindent i) the SU(2) symmetric phase, with no broken generator and all
the fields $A_{a\mu}$ massless, in which Eqs.(\ref{24}) is solved for ${\vec
\pi}_a$.

\noindent ii) 3 phases with SU(2) broken to two not-commuting U(1)'s [one
broken and two unbroken generators], in which Eqs.(\ref{24}) are solved in
two of the ${\vec \pi}_a$'s and one of the $\pi_{\theta a}$'s. The three phases
are: a) $A_{1\mu}, A_{2\mu}$ massless and $A_{3\mu}$ massive; b) $A_{3\mu},
A_{1\mu}$ massless and $A_{2\mu}$ massive; c) $A_{2\mu}, A_{3\mu}$ massless
and $A_{1\mu}$ massive. Naturally there are many more possibilities, because
one could choose any combination of the $A_{a\mu}$'s as the massive field.

\noindent iii) 3 phases with SU(2) broken to U(1) [two broken and one 
unbroken generator], in which Eqs.(\ref{24}) are solved in one of the ${\vec 
\pi}_a$'s and two of the $\pi_{\theta a}$'s. The three phases are: a) $A_{1\mu}$
massless and $A_{2\mu}, A_{3\mu}$ massive; b) $A_{3\mu}$ massless and $A_{1\mu},
A_{2\mu}$ massive; c) $A_{2\mu}$ massless and $A_{3\mu}, A_{1\mu}$ massive.
Again it is arbitrary which combination of the $A_{a\mu}$'s is chosen to
remain massless.

\noindent iv) the Higgs phase with SU(2) totally broken and all the $A_{a\mu}$ 
massive. Eqs.(\ref{24}) are solved in the $\pi_{\theta a}$'s.

Instead of going on with the search of Dirac's observables  with respect to
the first class constraints $\pi^o_a(x)\approx 0$, $\Gamma_a(x)\approx 0$
as in the first method of I, we shall use the easiest alternative path
corresponding to the second method of I. Since the Lagrangian density
${\cal L}(x)$  is invariant under gauge transformations, let us do the
field-dependent gauge transformation ${\tilde U}^{-1}_{\theta}(x)$ [like
for going to the unitary gauge] on Eq.(\ref{18}). The final result is

\begin{eqnarray}
{\cal L}^{'}(x)&=&-{1\over {4g^2}}F^{'}_{a\mu\nu}(x)F_a^{{'}\, \mu\nu}(x)+
\phi_o^2(1+{1\over {\sqrt{2}\phi_o
}}H(x))^2\sum_aA^{{'}2}_a(x)+\nonumber \\
&+&{1\over 2}\partial_{\mu}H(x)\partial^{\mu}H(x)-2(\phi_o
\sqrt{\lambda})^2H^2(x)(1+{1\over {2\sqrt{2}\phi_o }}H(x))^2
+\nonumber \\
&+&{i\over 2}{\bar \psi}^{'}(x)[\gamma^{\mu}(\partial_{\mu}+A^{'}_{a\mu}(x)
T^a)-\loarrow{(\partial_{\mu}-A^{'}_{a\mu}(x)T^a)}\gamma^{\mu}]\psi^{'}(x)
-m{\bar \psi}^{'}(x)\psi^{'}(x),
\label{25}
\end{eqnarray}

\noindent with the new Lagrangian density depending only on the configuration
variables

\begin{eqnarray}
&&A^{'}_{a\mu}(x)=(A^{U_{\theta}}_{\mu}(x))_a\nonumber \\
&&\psi^{'}(x)={\tilde U}^{-1}_{\theta}(x)\psi (x)\nonumber \\
&&H(x).
\label{26}
\end{eqnarray}

The new canonical momenta are

\begin{eqnarray}
&&\pi^{{'}0}_a(x)=0\nonumber \\
&&{\vec \pi}^{'}_a(x)=g^{-2}{\vec E}^{'}_a(x)\nonumber \\
&&\pi_H(x)=\partial^oH(x),
\label{27}
\end{eqnarray}

\noindent plus the fermion momenta. The resulting Dirac Hamiltonian density is

\begin{eqnarray}
{\cal H}^{'}_D(x)&=&{1\over 2}\sum_a[g^2{\vec \pi}^{{'}2}_a(x)+g^{-2}{\vec B}
^{{'}2}_a(x)]-\nonumber \\
&-&\phi_o^2(1+{1\over
{\sqrt{2}\phi_o}}H(x))^2\sum_a [A^{{'}2}
_{ao}(x)-{\vec A}^{{'}2}_a(x)]+\nonumber \\
&+&{1\over 2}[\pi_H^2(x)+[\vec \partial H(x)]^2]+2(\phi_o
\sqrt{\lambda})^2H^2(x)(1+{1\over {2\sqrt{2}\phi_o }}H(x))^2
+\nonumber \\
&+&{i\over 2}{\bar \psi}^{'}(x)[\vec \gamma \cdot (\vec \partial+{\vec A}^{'}
_{a}(x)T^a)-\loarrow{(\vec \partial-{\vec A}^{'}_{a}(x)T^a)}\cdot \vec
\gamma ]\psi^{'}(x)+m{\bar \psi}^{'}(x)\psi^{'}(x)-\nonumber \\
&-&A^{'}_{ao}(x)[-{\hat {\vec D}}^{(A^{'})}_{ab}\cdot {\vec \pi}^{'}_b(x)+
i\psi^{{'}\, \dagger}(x)T^a\psi^{'}(x)]+\lambda_{ao}(x)\pi^{{'}o}_a(x).
\label{28}
\end{eqnarray}

The time constancy of the primary constraints $\pi^{{'}o}_a(x)\approx 0$
generates the secondary constraints

\begin{equation}
\zeta_a(x)=2[\phi_o+{1\over {\sqrt{2}}}H(x)]^2A^{'}_{ao}(x)-\vec
\partial \cdot {\vec \pi}^{'}_a(x)-c_{abc}{\vec A}^{'}_b(x)\cdot {\vec \pi}
^{'}_c(x)+i\psi^{{'}\, \dagger}(x)T^a\psi^{'}(x)\approx 0.
\label{29}
\end{equation}

The time constancy of the onstraints $\zeta_a(x)\approx 0$ determines the three
Dirac multipliers $\lambda_{ao}(x)$. Therefore, we get three pairs of
second class constraints $\pi^{{'}o}_a(x)\approx 0$, $\zeta_a(x)\approx 0$,
eliminating the three pairs $A^{'}_{ao}(x), \pi^{{'}o}_a(x),$ of canonical
variables by going to Dirac brackets. The final canonical basis of Dirac's
observables is

\begin{equation}
{\vec A}^{'}_a(x),\quad {\vec \pi}_a^{'}(x)=g^{-2}{\vec E}^{'}_a(x),\quad
H(x),\quad \pi_H(x),\quad \psi^{'}_{A\alpha}(x),\quad {\bar \psi}^{'}
_{A\alpha}(x);
\label{30}
\end{equation}

\noindent the physical fields ${\vec A}^{'}_a, \psi^{'},$ have been dressed 
with a cloud of Higgs would-be Goldstone fields $\theta_a(x)$.

By going to Dirac brackets with respect to the second class constraints, we
find the physical reduced Hamiltonian density of the Higgs phase
[we rescale the fields: ${\vec A}^{'}_a\mapsto g{\vec A}_a, g{\vec \pi}^{'}_a=
g^{-1}{\vec E}^{'}_a\mapsto {\vec \pi}_a={\vec E}_a, \psi^{'}\mapsto \psi$ 
so that $F_{a\mu\nu}=\partial_{\mu}A_{a\nu}-\partial_{\nu}A_{a\mu}+gc_{abc}
A_{b\mu}A_{c\nu}$ and $A_{ao}\equiv {1\over 2}[\vec \partial \cdot {\vec \pi}_a+
c_{abc}{\vec A}_b\cdot{\vec \pi}_c-ig\psi^{\dagger}T^a\psi ]/g^2[\phi_o+
{1\over {\sqrt{2}}}H]^2$]

\begin{eqnarray}
{\cal H}^{(Higgs)}(x)&=&{1\over 2}\sum_a[{\vec \pi}^2_a(x)+{\vec B}
^2_a(x)]+{1\over 2}m^2_A(1+{|g|\over {m_A}}H(x))^2\sum_a{\vec A}^2_a(x)
+\nonumber \\
&+&{1\over 2}[\pi^2_H(x)+(\vec \partial H(x))^2]+{1\over 2}m^2_HH^2(x)(1+
{|g|\over {2m_A}}H(x))^2+\nonumber \\
&+&\sum_a
{ {[\vec \partial \cdot {\vec \pi}_a(x)+gc_{abc}{\vec A}_b(x)\cdot {\vec \pi}
_c(x)-ig\psi^{\dagger}(x)T^a\psi (x)]^2}\over {2m^2_A[1+{|g|\over {m_A}}
H(x)]^2}}+\nonumber \\
&+&{i\over 2}\bar \psi (x)[\vec \gamma \cdot (\vec \partial +g{\vec A}_a(x)T^a)-
\loarrow{(\vec \partial -g{\vec A}_a(x)T^a)}\cdot \vec \gamma ]\psi (x)+
m\bar \psi (x)\psi (x),
\label{31}
\end{eqnarray}

\noindent where the original parameters $\phi_o, \lambda$, have been replaced
by the masses of the gauge and residual Higgs fields

\begin{eqnarray}
&&m_A={\sqrt{2}}|g|\phi_o,\quad\quad m_H=2\phi_o\sqrt{\lambda},
\nonumber \\
&&\Rightarrow \phi_o={{m_A}\over {\sqrt{2}|g|}},\quad\quad \lambda ={{g^2m^2_H}
\over{2m^2_A}}.
\label{32}
\end{eqnarray}

We see that the Hamiltonian is local, because the Higgs mechanism produces a
local, even if not polynomial, self-energy term. However, this self-energy
term yields a non-local relation between $\partial^o\vec A$ and ${\vec \pi}_a$ 
(like in the Abelian case of I)

\begin{eqnarray}
\partial^oA^{i}_a(\vec x,x^o)&=&\lbrace A^{i}_a(\vec x,x^o),\int d^3y{\cal H}
^{(Higgs)}_{phys}(\vec y,x^o)\rbrace =\nonumber \\
&=&-\pi^i_a(\vec x,x^o)+\nonumber \\
&+&\partial^i\, {{\vec \partial \cdot {\vec \pi}_a 
(\vec x,x^o)+gc_{abc}{\vec A}_b(\vec x,x^o)\cdot {\vec \pi}_c(\vec x,x^o)-
ig\psi^{\dagger}(\vec x,x^o)T^a\psi (\vec x,x^o)}\over {(m_A+
|g|H(x))^2}}\nonumber \\
&&{}\nonumber \\
\Rightarrow &\pi^i_a&(x)=-\partial^oA^{i}_a(x)-\nonumber \\
&-&\partial^i\, {1\over {\triangle +(m_A+|g|H(x))^2}}\times 
\nonumber \\ 
&&[\vec \partial \cdot \partial^o{\vec A}_a(x)-gc_{abc}{\vec A}_b(x)\cdot
{\vec \pi}_c(x)+ig\psi^{\dagger}(x)T^a\psi (x)],\nonumber \\
\Rightarrow &[& Z^{(A)}_{ab}(\vec x,x^o) +(m_A+|g| H(\vec x,x^o))^2{{\delta
_{ab}}\over {\triangle}} ] \triangle {1\over {\triangle +(m_A+|g| H(\vec x,
x^o))^2}} \nonumber \\
&&(\partial^o \vec \partial \cdot {\vec A}_b(\vec x,x^o)-gc_{buv}{\vec A}_u
(\vec x,x^o)\cdot {\vec \pi}_v(\vec x,x^o) +ig\psi^{\dagger}(\vec x,x^o)T^b\psi
(\vec x,x^o))=\nonumber \\
&=& \partial^o \vec \partial \cdot {\vec A}_a(\vec x,x^o)+gc_{abc}{\vec A}_b
(\vec x,x^o)\cdot \partial^o {\vec A}_c(\vec x,x^o)+ig\psi^{\dagger}(\vec x,x^o)
T^a\psi (\vec x,x^o),
\label{33}
\end{eqnarray}

\noindent Here $Z^{(A)}_{ab}$ is the following operator [see Eq.(3-12) of
the second paper in Ref.\cite{lus} for the determination of its Green function 
$G^{(A,Z)}_{o,ab}$]

\begin{eqnarray}
Z^{(A)}_{ab}(\vec x,x^o)&=& \delta_{ab}+gc_{abc} {\vec A}_c(\vec x,x^o)\cdot
{{\vec \partial}\over {\triangle}},\nonumber \\
&&Z^{(A)}_{ab}(\vec x,x^o) G^{(A,Z)}_{o,bc}(\vec x,\vec y;x^o)=\delta_{ac}
\delta^3(\vec x-\vec y),\nonumber \\
&&G^{(A,Z)}_{o,ab}(\vec x,\vec y;x^o)=-{\vec \partial}_x\cdot {\vec \zeta}^{(A)}
_{ab}(\vec x,\vec y;x^o)=\nonumber \\
&&=-{\vec \partial}_x\cdot [\vec c(\vec x-\vec y) (P\, e^{\int_{\vec y}
^{\vec x}\, d\vec z\cdot {\vec A}_u(\vec z,x^o) {\hat T}^u} )_{ab}]
\label{z1}
\end{eqnarray}

\noindent with the path ordering along the flat geodesic and with $\vec c
(\vec x)={{\vec \partial}\over {\triangle}} \delta^3(\vec x)= {{\vec x}\over
{4\pi | \vec x |^3}}$.

To invert the operator $Z^{(A)}_{ab}+\delta_{ab}(m_A+|g| H(x))^2{1\over
{\triangle}}$, we need its Green function $G^{(A,Z)}_{ab}$, which is given by
(assuming that the perturbation of the Higgs field is small so that the series
converges)

\begin{eqnarray}
G^{(A,Z)}&=&G_o^{(A,Z)}-G_o^{(A,Z)}TG_o^{(A,Z)}+G_o^{(A,Z)}TG_o^{(A,Z)}TG_o
^{(A,Z)}-.....\nonumber \\
&{}&\nonumber \\
&&T_{ab}(\vec x,x^o)=(m_A+|g| H(\vec x,x^o))^2 {1\over {\triangle_{\vec x}}}.
\label{z2}
\end{eqnarray}

Finally we get

\begin{eqnarray}
\pi^i_a(x)&=&-\partial^o A^i_a(x)-\partial^i{1\over {\triangle_x}} \int d^3y
G^{(A,Z)}_{ab}(\vec x,\vec y;x^o)\nonumber \\
&&(\partial^o\vec \partial_y \cdot {\vec A}_a(\vec y,x^o)+gc_{abc}{\vec A}_b
(\vec y,x^o)\cdot \partial^o{\vec A}_c(\vec y,x^o)+ig\psi^{\dagger}(\vec y,x^o)
T^a\psi (\vec y,x^o),
\label{z3}
\end{eqnarray}

\noindent and a nonlocal Lagrangian density describing only the Higgs
phase

\begin{eqnarray}
{\cal L}^{(Higgs)}_{phys}(x)&=&\psi^{\dagger}(x)[i\partial^o-\vec
\alpha \cdot (i\vec \partial +e{\vec A}_a(x)T^a)-\beta m]\psi (x)-
\nonumber \\
&-&{1\over 2}[ {1\over {\triangle_x}} \int d^3y G^{(A,Z)}_{ab}(\vec x,\vec y;
x^o) ( \partial^o \vec \partial_y\cdot {\vec A}_b(\vec y,x^o)+\nonumber \\
&+&gc_{buv}{\vec A}_u(\vec y,x^o)\cdot \partial^o{\vec A}_v(\vec y,x^o)+ig
\psi^{\dagger}(\vec y,x^o)T^b\psi (\vec y,x^o) )]\times \nonumber \\
&&(\triangle +(m_A+|g| H(\vec x,x^o)) [ {1\over {\triangle_x}} \,
\int d^3z G^{(A,Z)}_{ac}(\vec x,\vec z;x^o) \nonumber \\
&&(\partial^o \vec \partial_z\cdot
{\vec A}_c(\vec z,x^o)+gc_{crs}{\vec A}_r(\vec z,x^o)\cdot \partial^o{\vec A}_s
(\vec z,x^o)+ig\psi^{\dagger}(\vec z,x^o)T^c\psi (\vec z,x^o) ) ]-\nonumber \\
&-&{1\over 2}m^2_A(1+{{|g|}\over {m_A}}H(x))^2\sum_a {\vec A}^{2}_a(x)
-{1\over 2}\sum_a{\vec B}^2_a(x)+\nonumber \\
&+&{1\over 2}\partial_{\mu}H(x)\partial^{\mu}H(x)-{1\over 2}
m^2_H H^2(x)(1+{{|g|}\over {2m_A}}H(x))^2.
\label{34}
\end{eqnarray}

Let us remark that in those points $x^{\mu}$ where $H(x)=-m_A/|g|=-
\sqrt{2}\phi_o$ [which were excluded to exist not to have problems
with the origin of the radial coordinates of Eq.(\ref{16})]
we would recover massless SU(2) gauge theory,
so that the numerator of the self-energy term in Eq.(\ref{32}) must vanish,
being the Gauss law of the massless theory. Therefore we should not have a
singularity in these points, but new physical effects like non-Abelian
vortices\cite{vort} in analogy to the Nielsen-Olesen vortices of the
Abelian case (see I); however now one needs to consider nontrivial SU(2)
bundles even in absence of non-Abelian monopoles\cite{go,pra}. See also
Ref.\cite{an}, where there is mass generation without the Higgs
mechanism from the requirement of integrability (absence of essential
singularities) of the equations of motion.

Let us remark that the self-energy appearing in Eq.(\ref{31}) is local and
that, in presence of fermion fields, it contains a 4 fermion interaction,
which has appeared from the nonperturbative solution of the Gauss laws and
which is a further obstruction to the renormalizability of the reduced theory
(equivalent to the unitary gauge, but without having added any gauge-fixing),
which already fails in the unitary physical gauge due to the massive vector
boson propagator not fulfilling the power counting rule; as said
in Ref.\cite{or}, this is due to the fact that the field-dependent gauge 
transformation relating $\vec A$ and ${\vec A}^{'}$ in Eq.(\ref{26}) is not
unitarily implementable. It is interesting to note that all the interaction
terms of the residual Higgs field $H(x)$ in Eq.(\ref{32}) show that it
couples to the ratio $|g|/m_A$.

As in the Abelian case of I, one can consistently eliminate\cite{kks}
the residual Higgs field $H(x)$ by adding with a multiplier the constraint 
$H(x)\approx 0$ at the physical Hamiltonian: its time constancy would produce 
the constraint $\pi_H(x)\approx 0$, and $\partial^o\pi_H(x)\approx 0$ would 
determine the multiplier; the final Hamiltonian density would be 

\begin{eqnarray}
{\cal H}^{(H=0)}(x)&=&{1\over 2}\sum_a[g^2{\vec \pi}^2_a(x)+g^{-2}{\vec B}
^2_a(x)]+{1\over 2}m^2_A\sum_a{\vec A}^2_a(x)+\nonumber \\
&+&{ {[\vec \partial \cdot {\vec \pi}_a(x)+c_{abc}{\vec A}_b(x)\cdot {\vec \pi}
_c(x)-ig\psi^{\dagger}(x)T^a\psi (x)]^2}\over {2m^2_A}}+\nonumber \\
&+&{i\over 2}\bar \psi (x)[\vec \gamma \cdot (\vec \partial +g{\vec A}_a(x)T^a)-
\loarrow{(\vec \partial -g{\vec A}_a(x)T^a)}\cdot \vec \gamma ]\psi (x)+
m\bar \psi (x)\psi (x),
\label{35}
\end{eqnarray}

The elimination of $H(x)$ reproduces the reduction to Dirac's observables of
the massive non-Abelian vector theory (like in the Abelian case of I) or Proca
field theory, whose Lagrangian density, in absence of fermions, is (see for
instance Ref.\cite{bs})

\begin{equation}
{\cal L}^{(mass)}(x)=-{1\over 4}F_{a\mu\nu}(x)F_a^{\mu\nu}(x)+{1\over 2}M^2
A_{a\mu}(x)A_a^{\mu}(x).
\label{36}
\end{equation}

The elimination of $H(x)$ can also be thought
as a limiting classical result of the so-called ``triviality problem"
[triviality of the $\lambda \phi^4$ theory \cite{tri}], which however would 
imply a quantization (but how?) of the
Higgs phase alone without the residual Higgs field, so that also its
quantum fluctuations would be absent. Instead these fluctuations are the main 
left quantum effect in the limit $m_H\rightarrow \infty$, which is known to
produce\cite{ab}, in the non-Abelian case, a gauge theory coupled to a
nonlinear $\sigma$-model, equivalent\cite{bs} to a massive Yang-Mills
theory. Indeed, in absence of fermions, the Lagrangian density (\ref{1}) can
be rewritten in terms of a linear $SU(2)_L\times SU(2)_R$ $\sigma$-model 
[$\lambda =m^2_H/4\phi_o^2$]

\begin{eqnarray}
{\tilde {\cal L}}(x)&=&-{1\over 4}F_{a\mu\nu}(x)F_a^{\mu\nu}(x)+{[D^{(A)}
_{\mu}\phi (x)]}^{\dagger}D^{(A)\mu}\phi (x)-{{m_H^2}\over {4\phi_o^2}}
[\phi^{\dagger}(x)\phi (x)-\phi_o^2]^2=\nonumber \\
&=&-{1\over 4}F_{a\mu\nu}(x)F_a^{\mu\nu}(x)+{1\over 2}Tr\, [D^{(A)}
_{\mu}M^{\dagger}(x)D^{(A)\mu}M(x)]-{{m^2_H}\over {4\phi_o^2}}[{1\over 2}
Tr\, M^{\dagger}(x)M(x) -\phi_o^2]^2\nonumber \\
&&{}\nonumber \\
M(x)&=&\left( \begin{array}{ll} \phi_1(x)&-\phi_2^{*}(x)\\ \phi_2(x) & 
\phi_1^{*}(x) \end{array} \right) =\sigma (x)+i\vec \tau \cdot \vec \xi (x).
\label{37}
\end{eqnarray}

For $m_H\rightarrow \infty$ one has $M^{\dagger}M\rightarrow \phi_o^2$
(i.e. $H(x)\rightarrow 0$; strongly interacting symmetry breaking sector), 
so that one gets $M(x)=\phi_oU(x)$ with 
$U^{\dagger}U=1$ and ${\tilde {\cal L}}\rightarrow -{1\over 4}F^2+
{{\phi_o^2}\over 2}Tr\, [D^{(A)}_{\mu}U^{\dagger}D^{(A)\mu}U]$, which is the
Lagrangian density of the nonlinear $SU(2)_L\times SU(2)_R$ $\sigma$-model
(broken to $SU(2)_{L+R}$ by $U^{\dagger}U=1$) with either $U(x)=e^{i\vec
\tau \cdot \vec \rho (x)}$ or $U(x)=\sqrt{1-{\vec \zeta}^2(x)}+i\vec \tau 
\cdot \vec \zeta (x)$.

As in I, one could add to either Eq.(\ref{1}) or Eq.(\ref{36}) a term
${1\over 2} \vec \partial A_{ao}(x) \cdot \vec \partial A_{ao}(x)$ [which
could be made Lorentz-covariant by reformulating the theory on spacelike
hypersurfaces\cite{lus,lus2,lv}], so that the local self-energy terms in
Eqs.(\ref{31}) and (\ref{35}) would be replaced by ${1\over 2}[\vec \partial 
\cdot {\vec \pi}_a(x)+c_{abc}{\vec A}_b(x)\cdot {\vec \pi}_c(x)-
ig\psi^{\dagger}(x)T^a\psi (x)] {1\over {\triangle +m_A^2(1+{{|g|}\over {m_A}}
H(x))^2}} [\vec \partial \cdot {\vec \pi}_a(x)+c_{abc}{\vec A}_b(x)\cdot {\vec 
\pi}_c(x)-ig\psi^{\dagger}(x)T^a\psi (x)]$ and ${1\over 2} [\vec \partial \cdot 
{\vec \pi}_a(x)+c_{abc}{\vec A}_b(x)\cdot {\vec \pi}_c(x)-ig\psi^{\dagger}(x)
T^a\psi (x)] {1\over {\triangle +m_A^2}} [\vec \partial \cdot {\vec \pi}_a(x)+
c_{abc}{\vec A}_b(x)\cdot {\vec \pi}_c(x)-ig\psi^{\dagger}(x)T^a\psi (x)]$,
respectively.

\section
{Comments}

i) As in the Abelian case of I, the covariant R-gauge-fixings\cite{rgau}

\begin{equation}
\partial^{\mu}A_{a\mu}(x)+\xi \theta_a(x)\approx 0,
\label{38}
\end{equation}

\noindent used in the proof of renormalizability and in the evaluation
of radiative corrections, are ambiguous like the Gauss laws: they can be solved 
either  in the Higgs fields (would-be Goldstone bosons) $\theta_a(x)$
[Higgs phase] or in $A_{ao}(x)$ [SU(2) phase] or in a mixed way [the other
four mixed phases]. It turns out that in the proofs of renormalizability
one is mixing all the existing disjoint phases (all of them are not
physical except the Higgs one; at most the SU(2) phase could be relevant in
cosmology, but not the mixed non-SU(2)-covariant ones), and only at the
end, in the limit $\xi \rightarrow \infty$, one is recovering the Higgs
phase.

ii) Let us now consider the non-Abelian SU(2) charges, whose existence is
implied by the Yang-Mills Noether identities\cite{lus,lus3}.

The gauge invariance of ${\cal L}(x)$ under the infinitesimal
gauge transformations $\delta A_{a\mu}(x)=\partial_{\mu}\alpha_a(x)+
c_{abc}A_{b\mu}(x)\alpha_c(x)={\hat D}^{(A)}_{\mu ab}\alpha_b(x)$,
$\delta \psi (x)=-\alpha_a(x)T^a \psi (x)$, $\delta \bar \psi (x)=\bar \psi (x)
\alpha_a(x)T^a$, $\delta \phi (x)=-i\alpha_a(x){\tilde T}^a \phi (x)$, 
$\delta \phi^{\dagger}=i \phi^{\dagger}\alpha_a(x){\tilde T}^a$,
produces the Noether identities

\begin{eqnarray}
0\equiv \delta {\cal L}&=&{{\partial {\cal L}}\over {\partial A_{a\mu}}}
\delta A_{a\mu}+{{\partial {\cal L}}\over {\partial \partial_{\nu}A_{a\mu}}}
\delta \partial_{\nu}A_{a\mu}+\delta \bar \psi {{\partial {\cal L}}\over
{\partial \bar \psi}}+\delta \partial_{\mu}\bar \psi {{\partial {\cal L}}
\over {\partial \partial_{\mu}\bar \psi}}+\nonumber \\
&+&\delta \psi {{\partial {\cal L}}\over
{\partial \psi}}+\delta \partial_{\mu}\psi {{\partial {\cal L}}
\over {\partial \partial_{\mu}\psi}}+{{\partial {\cal L}}\over {\partial \phi}}
\delta \phi +{{\partial {\cal L}}\over {\partial \partial_{\mu}\phi}}
\delta \partial_{\mu}\phi +\nonumber \\
&+&{{\partial {\cal L}}\over {\partial \phi^{\dagger}}}
\delta \phi^{\dagger} +{{\partial {\cal L}}\over {\partial \partial_{\mu}\phi
^{\dagger}}}\delta \partial_{\mu}\phi^{\dagger}=\nonumber \\
&=&g^{-2}L_a^{\mu}\delta A_{a\mu}+\delta \bar \psi L_{\bar \psi}-L_{\psi}\delta 
\psi +\delta \phi_i^{*}L_{\phi^{\dagger}\,i}+L_{\phi \, i}\delta \phi_i +
\partial_{\mu}G^{\mu}
\nonumber \\
&&{}\nonumber \\
G^{\mu}&=&\alpha_aG^{\mu}_{1a}+\partial_{\nu}\alpha_aG^{\mu\nu}_{oa}=
\nonumber \\
&=&-g^{-2}F_a^{\mu\nu}\delta A_{a\nu}-{i\over 2}[\delta \bar \psi \gamma^{\mu}
\psi -\bar \psi \gamma^{\mu}\delta \psi ]-{[D^{(A)\mu}\phi ]}^{\dagger}\delta 
\phi +\delta \phi^{\dagger} D^{(A)\mu}\phi \nonumber \\
&&\Downarrow \nonumber \\
G^{\mu\nu}_{oa}&=&-g^{-2}F_a^{\mu\nu}\nonumber \\
G^{\mu}_{1a}&=&-g^{-2}c_{abc}F^{\mu\nu}_bA_{c\nu}-i
\bar \psi \gamma^{\mu}T^a\psi +{[D^{(A)\mu}\phi ]}^{\dagger}{\tilde T}^a\phi -
\phi^{\dagger}{\tilde T}^aD^{(A)\mu}\phi=\nonumber \\
&=&-g^{-2}c_{abc}F_b^{\mu\nu}A_{c\nu}-J_a^{\mu}\nonumber \\
&&{}\nonumber \\
\partial_{\mu}G^{\mu}&=&\partial_{\mu}\partial_{\nu}\alpha_aG^{\mu\nu}_{oa}+
\partial_{\mu}\alpha_a[\partial_{\nu}G^{\nu\mu}_{oa}+G^{\mu}_{1a}]+\alpha_a
\partial_{\mu}G^{\mu}_{1a}\equiv \nonumber \\
&\equiv& -g^{-2}L_a^{\mu}\delta A_{a\mu}+L_{\psi}\delta \psi -\delta \bar \psi 
L_{\bar \psi}-L_{\phi \, i}\delta \phi_i -\delta \phi_i^{*}L_{\phi^{\dagger} 
\, i}{\buildrel \circ \over =} 0.
\label{39}
\end{eqnarray}

The last line implies the Noether identities [$(\mu\nu )$ and $[\mu\nu ]$ mean
symmetrization and antisymmetrization respectively]

\begin{eqnarray}
&&G^{(\mu\nu )}_{oa}\equiv 0\nonumber \\
&&\partial_{\nu}G^{\nu\mu}_{oa}\equiv -G^{\mu}_{1a}+g^{-2}L_a^{\mu}=g^{-2}
(L_a^{\mu}+c_{abc}F^{\mu\nu}_bA_{c\nu})+
i\bar \psi \gamma^{\mu}T^a\psi +\phi^{\dagger}{\tilde T}^aD^{(A)\mu}\phi -
[D^{(A)\mu}\phi ]^{\dagger}{\tilde T}^a\phi \nonumber \\
&&\partial_{\mu}G^{\mu}_{1a}\equiv g^{-2}c_{abc}A_{b\mu}L^{\mu}_c+
L_{\psi}T^a\psi +\bar \psi T^aL_{\bar \psi}+
L_{\phi \,i}{\tilde T}^a\phi_i -\phi_i^{*}{\tilde T}^aL_{\phi^{\dagger} \, i}
{\buildrel \circ \over =} 0
\label{40}
\end{eqnarray}

\noindent and, from the last two lines of these equations, the contracted
Bianchi identities

\begin{equation}
{\hat D}^{(A)}_{\mu}L^{\mu}-g^2{\hat T}^a(L_{\psi}T^a\psi +\bar \psi T^a
L_{\bar \psi}+L_{\phi \, i}{\tilde T}^a\phi_i-\phi_i^{*}{\tilde T}^a
L_{\phi^{\dagger} \, i})\equiv 0.
\label{41}
\end{equation}

The following subset of Noether identities reproduces the Hamiltonian
constraints

\begin{eqnarray}
&&\pi^o_a=G^{(oo)}_{oa}\equiv 0\nonumber \\
&&0\equiv \partial^o \pi_a^o\equiv -\Gamma_a -g^{-2}L_a^{o}
{\buildrel \circ \over =}-\Gamma_a.
\label{42}
\end{eqnarray}

The strong improper conservation laws\cite{lus3}
$\partial_{\mu}V_a^{\mu}\equiv 0$, implied
by Eqs.(\ref{40}), identify the strong improper conserved currents
(strong continuity equations)

\begin{eqnarray}
-V_a^{\mu}&=&\partial_{\nu}G^{\nu\mu}_{oa}=g^{-2}\partial_{\nu}F_a^{\nu\mu}=
\partial_{\nu}U_a^{[\mu\nu ]}\quad
{\buildrel \circ \over =}\quad g^{-2}c_{abc}F_b^{\mu\nu}A_{c\nu}+J_a^{\mu}= 
\nonumber \\
&=&g^{-2}c_{abc}F^{\mu\nu}_bA_{c\nu}+i\bar \psi \gamma^{\mu}T^a\psi +
[D^{(A)\mu}\phi]^{\dagger}{\tilde T}^a\phi -\phi^{\dagger}{\tilde T}^a
D^{(A)\mu}\phi  =\nonumber \\
&=&-G^{\mu}_{1a}=g^{-2}c_{abc}F_b^{\mu\nu}A_{c\nu}
+j_{F\, a}^{\mu}+j_{KG\, a}^{\mu},
\label{43}
\end{eqnarray}

\noindent with the superpotential $U_a^{[\mu\nu ]}=-g^{-2}F^{a\mu\nu}$. In 
the last line, $j^{\mu}_{F\, a}=i\bar \psi \gamma^{\mu}T^a\psi$ and 
$j_{KG\, a}^{\mu}=[D^{(A)\mu}\phi]^{\dagger}{\tilde T}^a\phi
-\phi^{\dagger}{\tilde T}^aD^{(A)\mu}\phi$ are the charge currents of the 
fermion field and of the complex Klein-Gordon Higgs fields respectively,
while $J^{\mu}_a$ is the total current of Eq.(\ref{6}).

The associated weak improper conservation laws are $\partial_{\mu}G^{\mu}_{1a}
{\buildrel \circ \over =} 0$ [it is obtained by using the second line of 
Eqs.(\ref{40})]. If $Q_a$ are the weak improper conserved 
non-Abelian Noether charges and
$Q^{(V)}_a$ the strong improper conserved ones, we get [its meaning is
equivalent to $\int d^3x \Gamma_a(\vec x,x^o){\buildrel \circ \over =} 0$]

\begin{eqnarray}
Q_a&=&-g^{-2}\int d^3x G^o_{1a}(\vec x,x^o)=-c_{abc}\int d^3x F^{ok}_b
(\vec x,x^o)A_c^k(\vec x,x^o)+g^2\int d^3x J_a^o(\vec x,x^o)=\nonumber \\
&=&\int d^3x[-c_{abc}{\vec A}_b(\vec x,x^o)\cdot {\vec E}_c(\vec x,x^o)+
ig^2\psi^{\dagger}(\vec x,x^o)T^a\psi (\vec x,x^o)-\nonumber \\
&-&g^2(\pi_{\phi}(\vec x,x^o){\tilde T}
^a\phi (\vec x,x^o)-\phi^{\dagger}(\vec x,x^o){\tilde T}^a
\pi_{\phi^{\dagger}}(\vec x,x^o))]=\nonumber \\
&=&g^2\int d^3x [-c_{abc}{\vec A}_b(\vec x,x^o)\cdot {\vec \pi}_c(\vec x,x^o)+
i\psi^{\dagger}(\vec x,x^o)T^a\psi (\vec x,x^o)+\nonumber \\
&+&\pi_{\theta b}(\vec x,x^o)V^{-1}_{ba}(\theta (\vec x,x^o))]=
\, Q_{F\, a}+Q_{\theta \, a}{\buildrel \circ \over =}\nonumber \\
{\buildrel \circ \over =} Q_a^{(V)}&=& g^2\int d^3x V^o_a(\vec x,x^o)=
\nonumber \\
&=&\int d^3x \partial^kF_a^{ko}(\vec x,x^o)=g^2\int d^3x
\vec \partial \cdot {\vec \pi}_a(\vec x,x^o)=\int d\vec \Sigma \cdot {\vec 
E}_a(\vec x,x^o),\nonumber \\
&&{}
\label{44}
\end{eqnarray}

\noindent where $Q_{F\, a}$ and $Q_{\theta \, a}$ are the non-Abelian charges 
(in units of $g$) of the fermion fields and of the complex Higgs field.
They are gauge covariant due to the assumed boundary conditions\cite{lus}.
For $\alpha =\alpha_a{\hat T}^a=const.$, one speaks of improper ``global"
[or ``rigid" or ``first kind"] gauge transformations [under them the gauge
potentials transform gauge covariantly\cite{lus}: $A\mapsto U^{-1}_{rigid}A
U_{rigid}$]; the generator of the infinitesimal improper gauge transformations
is $G[\alpha ]=-\int d^3x \alpha_a(\vec x,x^o)\Gamma_a(\vec x,x^o)=
-\int d^3x [{\vec \pi}_a(\vec x,x^o){\vec {\hat D}}^{(A)}_{ab}
\alpha_b(\vec x,x^o)-\vec \partial \cdot (\alpha_a(\vec x,x^o){\vec \pi}_a
(\vec x,x^o))]\, {=}_{\alpha_a=const.}\, -\int d^3x c_{abc}{\vec \pi}_a(\vec
x,x^o)\cdot {\vec A}_b\alpha_c+\sum_a \alpha_a Q^{(V)}_a$.

In the SU(2) phase, $Q_a{\buildrel \circ \over =} Q_a^{(V)}$ is the Gauss
theorem associated with the long-range massless SU(2) interaction: 
the flux at space infinity of the non-Abelian electric field is equal to the
total non-Abelian charge of the fermions and of the charged complex Higgs
fields, dressed with their Coulomb clouds, with the additional information 
that the Higgs non-Abelian charge is carried by the phases $\theta_a(x)$. 
The Dirac observables for the non-Abelian charges $Q_a$ could be evaluated
with the method of Ref.\cite{lus}.

On the contrary, in the Higgs phase $Q_a^{(V)}=0$ because the electric fields
${\vec E}_a$ decay at infinity due to the generated mass $m_A$ of the
SU(2) fields (the interactions have become short-range). Therefore, in presence
of spontaneous symmetry breaking through the Higgs mechanism, we loose the
Gauss theorem; Eq.(\ref{44}) only says that $Q_{F\, a}{\buildrel \circ \over =}
-Q_{\theta \, a}$, like in the Abelian case of I, before doing the
canonical reduction to the Dirac observables; but this is the statement
that each fermion and vector field is going to be dressed by a Higgs cloud of
would-be Goldstone bosons $\theta_a(x)$. In the Higgs sector, the original
SU(2) local gauge symmetry is reduced to a global one, under which the
Lagrangian density (\ref{34}) is invariant. This implies that the non-Abelian
charges $Q_{F\, a}$ become ordinary Noether constants in the Higgs phase.
Their expression in terms of Dirac observables (in units of $g$) is

\begin{eqnarray}
&&{\check Q}_{F\, a}=\int d^3x [c_{abc}{\vec A}^{'}_b(\vec x,x^o)\cdot
{\vec E}_c(\vec x,x^o)+i\psi^{{'}\dagger}(\vec x,x^o)T^a\psi^{'}(\vec x,x^o)]
\nonumber \\
&&\lbrace {\check Q}_{F\, a},{\check Q}_{F\, b}\rbrace =c_{abc}{\check Q}
_{F\, c}.
\label{45}
\end{eqnarray}

iii)
Let us remark that, since in the Higgs phase the Gauss law constraints are
solved algebraically in the Higgs momenta, we did not need to make
additional assumptions about the functional space the gauge potentials and gauge
transformations belong to as in Ref.\cite{lus} to avoid Gribov ambiguity.
However the assumptions of Ref.\cite{lus} are necessary to find the Dirac
observables of the other phases, because only in this way all forms of Gribov
ambiguity are avoided: in this way neither any gauge potential nor any field
strength has a nontrivial stability subgroup of gauge transformations 
[otherwise, for gauge potentials one has ``gauge symmetries" and the associated
stratification of the constraint manifold and of the reduced phase space,
while for field strengths one has the problem of ``gauge copies"]. In
particular, without these assumptions, there exist special gauge potentials
${\tilde A}_{a\mu}(x)$
with nontrivial (i.e. different from the center of the group of gauge
transformations) stability subgroup; the gauge transformations $\tilde U$ 
belonging to these subgroups are covariantly constant [${\hat D}^{(\tilde A)}
_{\mu}\tilde U=0$] and are called ``gauge symmetries" of these gauge potentials,
which correspond to reducible connections on the principal bundle of the
theory. For $G=SU(2)$ the only possible stability subgroups are U(1) and
$Z_2$; with each inequivalent $U(1)\subset SU(2)$ is associated a stratum of
reducible gauge potentials (connected by gauge transformations) with 
reduced structure group H=U(1) and with stability subgroup (or stabilizer) of
gauge transformations $Z_{SU(2)}[U(1)]=U(1)$ [$Z_G(H)=\lbrace a\in G\, |\, 
ab=ba\,\, for\, each\, b\in H\rbrace$ is the centralizer of H in G].
For $G\not= SU(2)$, if H is the reduced structure group of a stratum (H also
is the holonomy group of all thew connections in the stratum), then its gauge
potentials have $Z_G(H)$ as stability subgroup of gauge transformations with 
$H\not= Z_G(H)$ in general; the selection of one of these strata, when the 
function space allows their existence, is a kind of symmetry breaking since G 
(the structure group of the main stratum without gauge symmetries) is reduced 
to H; if also the field strength has a stability subgroup $G_F$ of gauge 
transformations  ($\hat U\in G_F$ iff $[\hat U,F_{a\mu\nu}{\hat T}^a]=0$),
one has $G_F\supseteq Z_G(H)$ with $H\subset G$ (and, if $\pi_1(G)\not= 0$,
each stratum has disjoined components) and there are ``gauge copies" $\hat A=
{\hat U}^{-1}A\hat U+{\hat U}^{-1}d\hat U$ of each gauge potential A in the
stratum with the same field strength $F_{a\mu\nu}[\hat A]=F_{a\mu\nu}[A]$.
Only in function spaces allowing the existence only of irreducible gauge
potentials [all of them have G as holonomy group], one has H=G and $G_F=Z_G(G)
=Z_G$ [$Z_G$ is the center of G] and there is no Gribov ambiguity.

As shown in Ref.\cite{he}, the existence of reducible gauge potentials has
implications for the non-Abelian charges associated with them. Let ${\tilde
A}_{\mu}={\tilde A}_{\tilde a\mu}{\hat T}^{\tilde a}$ be a gauge potential with
holonomy group $H\subset G$ [${\hat T}^{\tilde a}$ are the generators of the
Lie subalgebra $g_H$ of the Lie algebra $g$ of G] and stability subgroup
$Z_G(H)$. Since ${\tilde A}_{\mu}\in g_H$, one has that also the corresponding
Euler-Lagrange equations belong to $g_H$: $L^{\mu}_{\tilde a}{\hat T}^{\tilde a}
=({\hat D}^{(A)}_{\nu \tilde a\tilde b}{\tilde F}^{\nu\mu}_{\tilde b}+g^2J^{\mu}
_{\tilde a}){\hat T}^{\tilde a}\in g_H$. If $U=\openone +\alpha_a{\hat T}^a\in
Z_G(H)$ is an infinitesimal gauge symmetry of ${\tilde A}_{\mu}$, then
${\hat D}^{(A)}_{\mu ab}\alpha_b=0$. In the Noether identities of Eq.(\ref{39})
one has $G^{\mu\nu}_{oa}\equiv 0$, $G^{\mu}_{1a}{\hat T}^a=J^{\mu}_a{\hat T}^a
\in g_{Z_G(H)}$ [the Lie algebra of $Z_G(H)$], $\partial_{\mu}G^{\mu}=\partial
_{\mu}\alpha_aG^{\mu}_{1a}+\alpha_a\partial_{\mu}G^{\mu}_{1a}\equiv L_{\psi}
\delta \psi -\delta \bar \psi L_{\bar \psi}
-L_{\phi\, i}\delta \phi_i-\delta \phi_i^{*}L
_{\phi^{\dagger}\, i}$, so that Eqs.(\ref{40}) are replaced by

\begin{eqnarray}
&&G^{\mu}_{1a}{\hat T}^a\equiv J^{\mu}_a{\hat T}^a \in g_{Z_G(H)}\nonumber \\
&&\partial_{\mu}G^{\mu}_{1a}{\hat T}^a\equiv (L_{\psi}T^a\psi +\bar \psi T^a
L_{\bar \psi}+L_{\phi\, i}{\tilde T}^a\phi_i-\phi^{*}_i{\tilde T}^aL_{\phi^{*}\,
i}){\hat T}^a\, {\buildrel \circ \over =}\, 0.
\label{46}
\end{eqnarray}

\noindent There is only the weak improper conserved Noether charge 

\begin{equation}
Q_a{\hat T}^a= \int d^3x J^o_a(\vec x,x^o){\hat T}^a \in g_{Z_G(H)}.
\label{47}
\end{equation}

\noindent This charge has in general two components, $Q_a{\hat T}^a=Q_1+Q_2\in
g_{Z_G(H)}$, with $Q_1=Q_{1\tilde a}{\hat T}^{\tilde a}\in g_{Z_G(H)}\cap g_H$
and $Q_2=Q_{2a}{\hat T}^a$ in $g_{Z_G(H)}$ but not in $g_H$. While $Q_2$ is
only a Noether charge  determined by the matter fields [a flavourlike charge
in the terminology of Ref.\cite{he}], there is another form associated with
$Q_1$, because one can use in it the $g_H$-valued Euler-Lagrange equations
$L^{\mu}_{\tilde a}{\hat T}^{\tilde a}{\buildrel \circ \over =} 0$:
$Q_2=\int d^3x J^o_{\tilde a}{\hat T}^{\tilde a}\, {\buildrel \circ \over =}\,
-g^{-2}\int {\hat D}^{(A)}_{\nu \tilde a\tilde b}{\tilde F}^{\nu o}_{\tilde b}
{\hat T}^{\tilde a}$. Therefore, $Q_1$ may be reexpressed only in terms of
the Yang-Mills field [itis a dynamical charge in the terminology of 
Ref.\cite{he}] as in the standard case of the Gauss theorem, and it can be 
shown\cite{he} that the charges $Q_{1\tilde a}$ are U(1)-charges corresponding
to the Abelian part of $Z_G(H)$; instead the charges $Q_{2a}$ are non-Abelian.

In our case with G=SU(2), the only possibility is $H=Z_G(H)=U(1)$ and there is
only an Abelian charge $Q_1$ associated with reducible gauge potentials.

iv) As said in I, to recover Lorentz covariance one has to reformulate the
theory on spacelike hypersurfaces as shown in Refs.\cite{lus,lus2,dir,lv}.
This reformulation produces an ultraviolet cutoff, because a classical unit 
of length $\rho =\sqrt{-W^2}/P^2$ emerges [the space domain over which
the noncovariance of the canonical center of mass of the field 
configuration extends], when we restrict to massive Poincar\'e
representations, $P^2 > 0$, $W^2\not= 0$. However, it is not yet clear how
to use this physical cutoff in the quantization of nonlocal and
nonpolynomial theories.

v) In a future paper we will unify the results of this paper and of
Refs.\cite{lus,lv} to determine the Dirac observables of the standard
$SU(3)\times SU(2)\times U(1)$ model.

\vfill\eject


\begin{references}

\bibitem{lv}L.Lusanna and P.Valtancoli, ``Dirac's Observables for the Higgs
model: I) the Abelian Case", Firenze University preprint, February 1996.
\bibitem{eb}P.W.Higgs, Phys.Rev.Lett. {\bf 13}, 508 (1964); Phys.Rev. 
{\bf 145}, 1156 (1966).
F.Englert and R.Brout, Phys.Rev.Lett. {\bf 13}, 321 (1964).
G.S.Guralnik, C.R.Hagen and T.W.B.Kibble, Phys.Rev.Lett. {\bf 13}, 585 (1964).
T.Kibble, Phys.Rev. {\bf 155}, 1554 (1967).
\bibitem{cl}T.P.Cheng and L.F.Li, ``Gauge Theory of Elementary Particle 
Physics", (Oxford University Press, New York, 1984).
\bibitem{lus}L.Lusanna, Int.J.Mod.Phys. {\bf A10}, 3531 and 3675 (1995).
\bibitem{lus1}L.Lusanna, ``Hamiltonian Constraints and Dirac's Observables"",
in ``Geometry of Constrained Dynamical Systems", Cambridge 1994, ed.J.M.Charap,
(Cambridge University Press, Cambridge, 1995).
\bibitem{go}P.Goddard and D.I.Olive, Rep.Prog.Phys. {\bf 41}, 1357 (1978).
\bibitem{sho}G.M.Shore, Ann.Phys.(N.Y.) {\bf 137}, 262 (1981).
\bibitem{fulp}R.O.Fulp and L.K.Norris, J.Math.Phys. {\bf 24}, 1871 (1983).
\bibitem{kk}Y.Kerbrat and H.Kerbrat-Lung, J.Geom.Phys. {\bf 3}, 221 (1986).
\bibitem{tra}A.Trautman, Czech.J.Phys. {\bf B29}, 107 (1979).
\bibitem{kn}S.Kobayasi and K.Nomizu, ``Foundations of Differential Geometry", 
Vol. I (Interscience, NewYork and London, 1963).
\bibitem{isham} C.J.Isham, J.Phys. {\bf A14}, 2943 (1981).
\bibitem{or}L.O'Raifeartaigh, ``Hidden Gauge Symmetries", Rep.Prog.Phys. 
{\bf 42}, 159 (1979).
\bibitem{str}F.Strocchi, ``Gauss'Law in Local Quantum Field Theory", in
``Field Theory, Quantization and Statistical Physics", ed.E.Tirapegui 
(Reidel, Dordrecht, 1981).
G.Morchio and F.Strocchi, in ``Fundamental Problems of Gauge Field Theory",
eds. G.Velo and A.S.Wightman, NATO ASI 141B (Plenum, New York, 1986).
\bibitem{lus2}L.Lusanna, ``N- and 1-time Classical Description of
N-body Relativistic Kinematics and the Electromagnetic Interaction",
Firenze Univ. preprint, January 1996.
\bibitem{dir}P.A.M.Dirac, ``Lectures on Quantum Mechanics", Belfer Graduate
School of Science, Monographs Series, (Yeshiva University, New York N.Y. 1964).
\bibitem{cosmo}L.G.Yaffe, ``The Electroweak Phase Transition: a Status Report",
talk at ``String Gravity and Physics at the Planck Energy Scale", Erice 1995.
\bibitem{vort}J.M.Cornwall, Nucl.Phys. {\bf B157}, 392 (1979).
\bibitem{pra}M.F.Atiyah and R.S.Ward, Commun.Math.Phys. {\bf 55}, 117 (1977).
R.Ward, Commun.Math.Phys. {\bf 79}, 317 (1981).
M.K.Prasad and P.Rossi, Phys.Rev. {\bf D24}, 2182 (1981).
\bibitem{an}J.T.Anderson, Mod.Phys.Lett. {\bf A3}, 1629 (1988); Int.J.Mod.
Phys. {\bf A7}, 201 (1992).
\bibitem{kks}Y.Kebrat, H.Kebrat-Lung and J.\'Sniatycki, Rep.Math.Phys.
{\bf 28}, 201 (1989).
\bibitem{bs}W.A.Bardeen and K.Shizuya, Phys.Rev. {\bf D18}, 1969 (1978).
\bibitem{tri}K.Wilson, Phys.Rev. {\bf B4}, 3184 (1971).
K.Wilson and J.Kogut, Phys.Rep. {\bf 12}, 75 (1974).
J.Fr\"olich, in ``Progress in Gauge Field Theory", Carg\`ese 1983, eds.
G.'t Hooft, A.Jaffe, H.Lehmann, P.K.Mitter, I.M.Singer and R.Stora,
NATO ASI B115 (Plenum, New York, 1984).
M.A.B.B\'eg and R.C.Furlong, Phys.Rev. {\bf D31}, 1370 (1985). 
\bibitem{ab}T.Appelquist and C.Bernard, Phys.Rev. {\bf D22}, 200 (1980).
\bibitem{rgau}B.W.Lee and J.Zinn-Justin, Phys.Rev. {\bf D5}, 3121, 3137 and 
3155 (1972).
A.Salam and J.Strathdee, Nuovo Cim. {\bf 11A}, 397 (1972).
K.Fujikawa, B.W.Lee and A.I.Sanda, Phys.Rev. {\bf D6}, 2923 (1972).
Y.P.Yao, Phys.Rev. {\bf D7}, 1647 (1973).
E.Abers and B.Lee, Phys.Rep. {\bf 9}, 1 (1975).
\bibitem{lus3}L.Lusanna, Riv.Nuovo Cimento {\bf 14} (3), 1 (1991).
\bibitem{he}A.Heil, A.Kersch, N.A.Papadopolous and B.Reifenh\"auser, Ann.Phys.
(N.Y.) {\bf 217}, 173 (1991).


\end{references}
\end{document}